\documentclass[aps,prl,twocolumn,amsmath,amssymb,10pt,aps,longbibliography,superscriptaddress,citeautoscript,bibnotes,floatfix,nobibnotes]{revtex4-2}

\usepackage[utf8]{inputenc}
\usepackage[T1]{fontenc}
\usepackage{textcomp} 

\usepackage{amsmath}   
\usepackage{amssymb}   
\usepackage{physics}   
\usepackage{bm}        
\usepackage{mathrsfs}  
\usepackage{mathtools} 

\usepackage{caption}  
\usepackage{graphicx} 
\graphicspath{ {./Figures/} }

\usepackage{subfigure} 

\usepackage{tabularx} 
\usepackage{booktabs} 
\usepackage{multirow} 
\usepackage{dcolumn}  

\usepackage[colorlinks=true,linkcolor=blue,citecolor=blue]{hyperref} 

\usepackage[normalem]{ulem} 
\usepackage{soul} 
\usepackage{amsthm}  
\usepackage{color,xcolor} 
\usepackage{lipsum}

\usepackage{titlesec}
\titleformat{\section}{\normalfont\large\bfseries}{}{}{}
\titlespacing{\section}{0pt}{10pt plus 2pt minus 2pt}{0pt}

\titleformat{\paragraph}{\normalfont\bfseries}{}{}{}
\titlespacing{\paragraph}{0pt}{10pt plus 2pt minus 2pt}{0pt}

\usepackage{comment}

\usepackage{soul}

\begin{document}
\title{Three-Dimensional Non-Foliated Fractional Quantum Hall Phases with Irrational Anyons in Twisted van der Waals Multilayers}

\author{Seyoung Jin}
\affiliation{Department of Physics, Pohang University of Science and Technology, Pohang, 37673, Republic of Korea}

\author{Hyeonseo Lim}
\affiliation{Department of Physics, Pohang University of Science and Technology, Pohang, 37673, Republic of Korea}

\author{Youngwook Kim}
\thanks{y.kim@dgist.ac.kr}
\affiliation{Department of Physics and Chemistry, Daegu Gyeongbuk Institute of Science and Technology (DGIST), Daegu 42988, Republic of Korea}

\author{Gil Young Cho}
\thanks{gilyoungcho@kaist.ac.kr}
\affiliation{Department of Physics, Korea Advanced Institute of Science and Technology, Daejeon 34141, Republic of Korea}
\affiliation{Center for Artificial Low Dimensional Electronic Systems, Institute for Basic Science, Pohang 37673, Korea}
 
\maketitle 

\textbf{Three-dimensional fractional quantum Hall phases offer a route to intrinsically higher-dimensional topological order beyond simple stacks of two-dimensional quantum Hall liquids. Such phases can exhibit non-foliated, intrinsically three-dimensional entanglement structures, exponentially large topological degeneracies and quasiparticles with irrational braiding statistics. Their microscopic realization has remained elusive because Landau quantization in three dimensions generally leaves dispersive one-dimensional bands, favoring metallic and density-wave states over incompressible fractional liquids. Here we show that large-angle twisted van der Waals multilayers provide a practical route around this obstruction. Large twist angles suppress coherent interlayer tunneling through momentum mismatch, while the atomic-scale layer separation preserves strong interlayer Coulomb interactions. Using Monte Carlo calculations to compare the energies of an extensive set of 862 competing trial wavefunctions, we find that generalized Halperin states with quantum coherence extending across multiple consecutive layers are stabilized. In experimentally accessible magnetic-field regimes, these states replace the metallic spontaneous-interlayer-coherent phases that dominate conventional untwisted graphite-like multilayers. The resulting liquids realize non-foliated fractional quantum Hall order closely related to fractonic topological order, hosting quasiparticles with rational electric charges but irrational braiding statistics. Their large topological degeneracy and non-rational statistical phases may offer unconventional resources for quantum information storage and processing. Our results establish twisted van der Waals multilayers as a realistic materials platform for three-dimensional fractional Hall matter beyond conventional two-dimensional quantum Hall systems.}

\section{Introduction} \noindent
The quantum Hall effect provides one of the clearest demonstrations of how topology, interactions and magnetic fields can reorganize electronic matter into phases with precisely quantized observables~\cite{NatRevPhys2020_vonKlitzing}. In two dimensions, the integer and fractional quantum Hall effects revealed dissipationless boundary transport, quantized Hall responses, fractionally charged quasiparticles and anyonic exchange statistics~\cite{PRB1982_Halperin, PRL1984_Halperin}. Extending this physics to three dimensions has been a long-standing goal~\cite{JJAPS1987_Halperin, PRB1992_Kohmoto, NatPhys2026_Seo}—not merely to construct a higher-dimensional analogue of the two-dimensional problem, but to realize intrinsically three-dimensional fractionalized matter. In particular, a three-dimensional fractional quantum Hall phase can possess non-foliated entanglement whose topological structure cannot be decomposed into a stack of independent two-dimensional quantum Hall layers~\cite{PRB2022_Ma}. Such phases are particularly intriguing because their topological degeneracy can grow exponentially with system thickness, suggesting applications to quantum information storage~\cite{RevModPhys2016_Brown,MPA2020_Pretko}. Moreover, their irrational braiding statistics may enable a dense set of topologically protected relative-phase rotations, offering a potential route towards phase gates~\cite{PRL2000_Naud,NPB2001_Sondhi,PRB2022_Ma,QPL2009_Wootton}. Three-dimensional quantum Hall physics, however, remains challenging to realize~\cite{PRL2020_Qin, npjQM2021_Li, RepProgPhys2023_Gooth}. Even the integer quantum Hall effect in three dimensions arises only under special circumstances~\cite{Nature2019_Tang, NatPhys2019_Yin}, and no unambiguous three-dimensional fractional Hall plateau has yet provided a comparable experimental benchmark~\cite{RepProgPhys2023_Gooth}. Identifying a realistic route to a stable three-dimensional fractional quantum Hall phase therefore remains a central challenge.

\begin{figure*}[t]
\begin{center}
\includegraphics[width=1\textwidth]{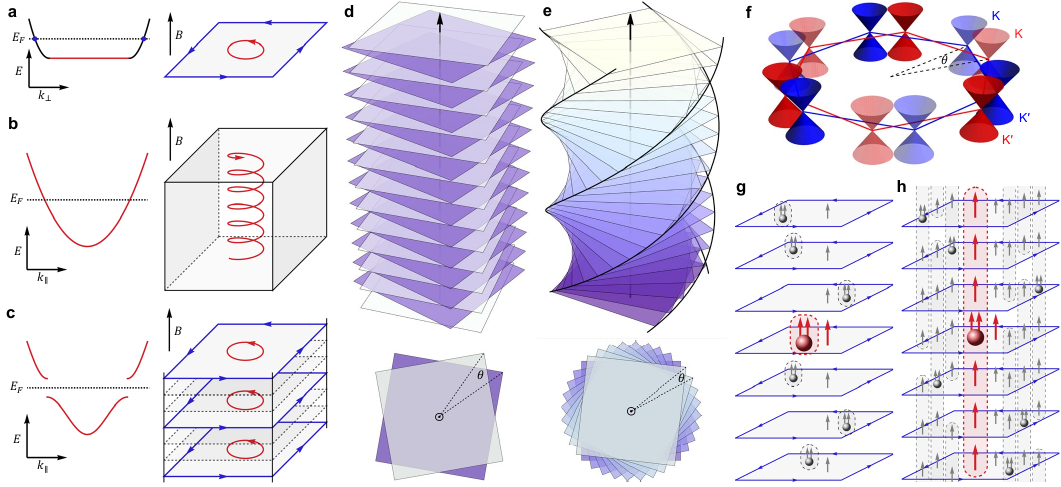}
\caption{
{\bf Three-dimensional FQH phases in twisted multilayers.} {\bf a,} Quantum Hall effect in two dimensions. Under a uniform magnetic field \(B\), the electrons follow cyclotron orbits (red arrow), and discrete Landau levels are formed~\cite{NatRevPhys2020_vonKlitzing, NatPhys2026_Seo}. An isolated flat Landau level is required to realize a quantum Hall state (red line). {\bf b,} In three dimensions, Landau levels generally disperse with the momentum \(k_\parallel\) along the applied magnetic field, corresponding to cyclotron motion that spirals along the field direction (red arrow)~\cite{NatPhys2026_Seo, NatPhys2019_Yin}. When a single Landau level crosses the Fermi energy (red curve), it results in a quasi-one-dimensional Fermi sea. {\bf c,} When this band is partially filled, it has a density-wave instability (right). {\bf d,} Alternating and {\bf e,} helical twisted multilayers with twist angle \(\theta\). {\bf f,} Large twist angle leads to a large momentum mismatch between neighboring layers (red and blue, respectively). {\bf g,} Stacked Laughlin state in composite fermion description~\cite{NatPhys2019_Csathy}, whose composite fermion consists of 2 intralayer vortices. Electrons and flux quanta are represented by spheres and arrows. {\bf h,} Generalized Halperin state \((3111)\), whose composite fermion consists of 2 intralayer and 6 interlayer vortices, distributed up to third-nearest neighbors.}
\label{fig1}
\end{center}
\end{figure*}

The difficulty originates from {the structure of Landau quantization} in three dimensions~\cite{PRL2020_Qin, npjQM2021_Li, RepProgPhys2023_Gooth}. In a two-dimensional electron system, a perpendicular magnetic field quenches the kinetic energy into discrete Landau levels separated by spectral gaps, allowing interactions within a partially filled, isolated Landau level to produce incompressible fractional quantum Hall liquids. In a three-dimensional metal, by contrast, electrons remain free to propagate along the magnetic-field direction. Landau levels consequently broaden into one-dimensional dispersive bands that typically cross the Fermi energy, forming quasi-one-dimensional Fermi seas rather than isolated flat bands (Fig.~\ref{fig1}a,b). This structure makes low filling factors difficult to access and often leaves several Landau bands simultaneously active near the Fermi energy. Moreover, even when a partially filled Landau band can be isolated, its residual one-dimensional character tends to favor density-wave instabilities over incompressible fractionalized liquids (Fig.~\ref{fig1}c). At the opposite extreme, a stack of decoupled two-dimensional layers may host fractional Hall liquids within each layer, but the resulting topological order is foliated—that is, it consists of a simple stack of two-dimensional states—rather than intrinsically three-dimensional. A viable platform must therefore meet a delicate requirement: coherent single-particle motion along the stacking direction must be suppressed, while sufficiently strong interlayer interactions are retained to entangle many layers.

Large-angle twisted van der Waals multilayers naturally satisfy these competing requirements. Twisted van der Waals materials have emerged as versatile platforms for correlated and topological phases, including superconductivity, correlated insulating states, strange metallicity and fractional quantum Hall phenomena in bilayer and few-layer systems~\cite{Nature2018_Cao_SC, Nature2018_Cao_Corr, PRL2020_Cao, Nature2023_Cai, Nature2023_Park, NanoLett2023_Kim, NatComm2025_Kim, NanoLett2026_Kim}. Recent experimental advances have further brought genuinely three-dimensional twisted structures within reach, including helical~\cite{Science2020_Zhao, Nature2024_Ji} and alternating twisted multilayers~\cite{NatMat2024_Wang, NatComm2025_Zhang} composed of many consecutively rotated layers  (Fig.~\ref{fig1}d,e). At large twist angles, the substantial momentum mismatch between low-energy states in adjacent layers strongly suppresses direct interlayer tunneling~\cite{PRL2013_Kim, NanoLett2023_Kim, NatComm2025_Kim} (Fig.~\ref{fig1}f). The interlayer Coulomb interaction, by contrast, remains strong because neighbouring layers are separated only by atomic-scale distances and can dominate over the residual tunneling under experimentally relevant magnetic fields~\cite{NanoLett2023_Kim, NatComm2025_Kim}. The low carrier densities and valley-resolved low-energy band structures of van der Waals materials further reduce the number of Landau levels active near the Fermi energy. Together, these features create a regime in which many quantum Hall layers are coupled predominantly through interactions rather than coherent single-particle hopping.

Here we show that large-angle twisted van der Waals multilayers provide a realistic route to non-foliated three-dimensional fractional quantum Hall phases. We perform Monte Carlo calculations of 862 competing trial wavefunctions to compare the energies between generalized Halperin liquids~\cite{PRB1990_Qiu, PRL2000_Naud, NPB2001_Sondhi, PRB2009_Burnell}, generalized composite-fermion crystals~\cite{PRL2013_Archer, PRB2020_Faugno}, staged liquid and crystalline states~\cite{PRB1988_MacDonald, PRB1990_Qiu, PRB2009_Burnell}, metallic spontaneous-interlayer-coherent states~\cite{PRB2009_Burnell, PRB2002_Hanna, PRB2003_Schliemann}, and their product-state and particle–hole-conjugate descendants. Comparing conventional three-dimensional Bernal-stacked graphite with alternating and helical twisted multilayers, we find that untwisted systems are dominated by metallic interlayer-coherent phases under experimentally accessible magnetic fields. Large-angle twisting suppresses these metallic states and instead stabilizes generalized Halperin liquids. These phases emerge prominently at fractional fillings such as $\nu_L = 1/7$ and $1/9$, and can remain stable at comparatively large filling fractions under magnetic fields as low as a few tesla. The resulting states exhibit quantum coherence across multiple consecutive layers, realize non-foliated three-dimensional topological order and host quasiparticles with rational electric charges but irrational braiding statistics~\cite{PRL2000_Naud, NPB2001_Sondhi, PRB2022_Ma}. Our results establish twisted van der Waals multilayers as an experimentally realistic platform for intrinsically three-dimensional fractional quantum Hall phases and irrational anyons in electronic materials.

\section{Results}
\paragraph{The Model.}\noindent
Our primary target system is three-dimensional twisted multilayer graphene, although the main results extend straightforwardly to twisted transition metal dichalcogenides. The effective Hamiltonian for twisted multilayer graphene projected onto the zeroth Landau level is 
\begin{align}\label{eq:Hamiltonian}
H_{\mathrm{eff}} = H_{\mathrm{Coulomb}} + H_{V} + H_{\mathrm{Tunnel}}, 
\end{align}
where $H_{\mathrm{Coulomb}}$ describes the long-range Coulomb interaction, $H_V$ denotes phenomenological short-range interactions~\cite{NatComm2017_Hunt, NanoLett2023_Kim, NatComm2025_Kim, NanoLett2026_Kim, PRB2006_Alicea, PRB2012_Kharitonov, PRL2012_Kharitonov}, and $H_{\mathrm{Tunnel}}$ represents electron tunneling between neighboring layers:
\begin{align*}
H_{\mathrm{Tunnel}}
=
-t_{\perp}
\sum_{l,\sigma}
\int d^2 \bm{r}\,
\psi^{\dagger}_{l,\sigma}(\bm{r})
\psi_{l+1,\sigma}(\bm{r})
+ \mathrm{h.c.}. \nonumber 
\end{align*}
Here, $\psi^{\dagger}_{l,\sigma}$ ($\psi_{l,\sigma}$) creates (annihilates) an electron in layer $l$ and valley $\sigma = K, K'$. We assume a sufficiently strong magnetic field such that the spin degeneracy is completely lifted. The effect of the large-angle twist is modeled as a strong suppression of the interlayer tunneling, e.g., $t_\perp \approx 1.78\,\mathrm{meV}$ in alternating twisted multilayer graphene with $\theta \gtrsim  10^{\circ}$~\cite{PRL2024_Lu, PRL2007_Santos}. In contrast, the conventional three-dimensional limit like graphite corresponds to much larger interlayer tunneling, $t_\perp \approx 10 \,\mathrm{meV}$~\cite{PRB2009_Burnell, PRL2007_Bernevig}. This separation of energy scales dramatically reshapes the phase diagram~(Fig.~\ref{fig3}), particularly in the experimentally accessible regime of magnetic fields. Some details of the Hamiltonian are presented in {Methods}. 

\begin{figure}[t]
\begin{center}
\includegraphics[width=1\columnwidth]{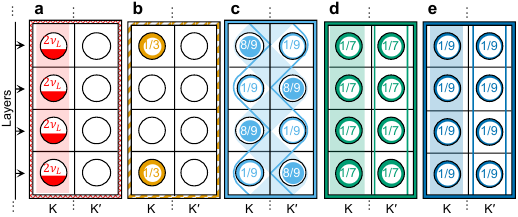}
\caption{
{\bf Schematics of competing phases.} Each black circle visualizes the lowest Landau level assigned to each layer (rows) and valley (columns). The circle is empty if the corresponding Landau level is unoccupied. Occupied Landau levels are indicated by partially filled circles, and their filling fractions are explicitly provided. A set of entangled Landau levels is indicated by either a shaded area or a solid line. {\bf a,} Metallic valley-polarized SILC state. {\bf b,} Staged Laughlin state. {\bf c--e,} Interlayer coherent phases as a pair of generalized Halperin states and/or their particle-hole conjugates. Using the notation in which \(\cdot\oplus\cdot\) denotes a product state and \(\mathrm{PH}(\cdot)\) denotes particle-hole conjugation, the representative generalized Halperin states are \((3111)\oplus\mathrm{PH}(3111)\) with opposite layer-alternating valley textures in {\bf c}, \((3110)\oplus(3110)\) with opposite polarized valley textures in {\bf d}, and \((3111)\oplus(3111)\) with opposite polarized valley textures in {\bf e}.}
\label{fig2}
\end{center}
\end{figure}

\begin{figure*}[t]
\begin{center}
\includegraphics[width=1\textwidth]{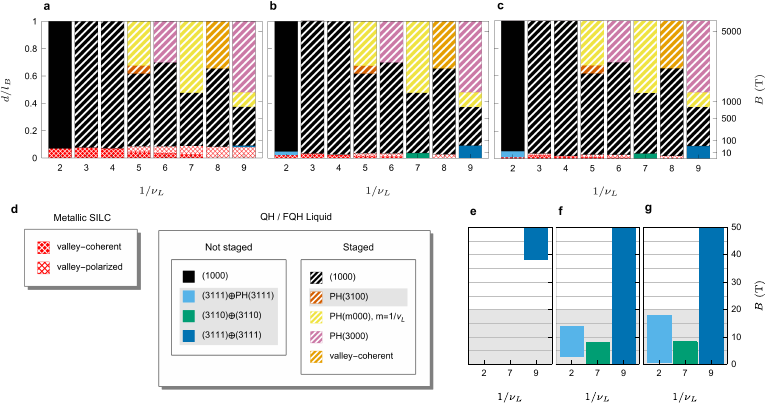}
\caption{
{\bf Phase diagrams.} {\bf a,} Phase diagram for three-dimensional Bernal graphite (\(t_\perp \approx 10 \text{ meV}\)), {\bf b,} alternating twisted graphene multilayer (\(t_\perp \approx 1.78 \text{ meV}\)), and {\bf c,} helical twisted graphene multilayer (\(t_\perp \approx 1.09 \text{ meV}\)). {\bf d,} Phase labels using the notation of Fig.~\ref{fig2}. Interlayer-coherent FQH phases are highlighted in gray. Valley ordering and entanglement patterns are provided in Supplementary Note 3. {\bf e--g,} Parameter regimes in which generalized Halperin states emerge at low magnetic fields for {\bf e,} Bernal graphite, {\bf f,} alternating twisted graphene multilayer, and {\bf g,} helical twisted graphene multilayer. Experimentally accessible range of magnetic fields is highlighted in gray.} 
\label{fig3}
\end{center}
\end{figure*}

\paragraph{Competing States.}\noindent
We consider a total of 862 wavefunctions describing competing liquid and crystalline states in the fractionally filled lowest Landau level as candidate ground states (Methods). They are broadly categorized into generalized Halperin states~\cite{PRB1990_Qiu, PRL2000_Naud, NPB2001_Sondhi, PRB2009_Burnell}, generalized composite-fermion crystals~\cite{PRL2013_Archer, PRB2020_Faugno}, staged liquid and crystalline states~\cite{PRB1988_MacDonald, PRB1990_Qiu, PRB2009_Burnell}, metallic spontaneous interlayer coherent (SILC) states~\cite{PRB2009_Burnell, PRB2002_Hanna, PRB2003_Schliemann}, and their product and/or particle-hole conjugate states, with possible intervalley and interlayer coherence. The explicit wavefunctions are provided in Supplementary Note 2.

The most intriguing states are generalized Halperin states, exhibiting multilayer entanglement and intrinsically three-dimensional fractionalization. The wavefunction of the state \((mnop)\) takes the form (Fig.~\ref{fig1}g,h)
\begin{align}\label{gHalperin}
&\Psi
\sim
\prod_l
\Biggl(
\prod_{i<j}(z_{l,\sigma_l, i}-z_{l,\sigma_l,j})^{m}
\prod_{i,j}(z_{l,\sigma_l,i}-z_{l+1,\sigma_{l+1},j})^{n}\nonumber\\
&\prod_{i,j}(z_{l,\sigma_l,i}-z_{l+2,\sigma_{l+2},j})^{o}
\prod_{i,j}(z_{l,\sigma_l,i}-z_{l+3,\sigma_{l+3},j})^{p}
\Biggr), 
\end{align}
up to the Gaussian factors. Here, $z_{l,\sigma,i}=x_{l,\sigma,i}+iy_{l,\sigma,i}$ denotes the holomorphic coordinate of the $i$-th electron in valley $\sigma$ and layer $l$. Since interlayer coherence are expected to decrease with increasing layer separation, we impose the hierarchy $m \geq n \geq o \geq p$, with $m \in 2\mathbb{Z}+1$ required by fermionic statistics of electrons. We note that this state can be naturally generalized to include intralayer intervalley coherence coexisting with interlayer coherence (Supplementary Note 2.1), and such generalized states are also incorporated into our set of candidate ground states. Representative states appearing in the phase diagram (Fig.~\ref{fig3}) are illustrated in Fig.~\ref{fig2}e.
  
A primary competing phase against the generalized Halperin states is the metallic SILC states favored by interlayer tunneling (Fig.~\ref{fig2}a). Intuitively, when the interlayer tunneling becomes sufficiently strong, electrons tend to delocalize along the magnetic-field direction and form a quasi-one-dimensional Fermi liquid (Fig.~\ref{fig1}b), resulting in SILC states. The corresponding wavefunction is transparently expressed in the second-quantized form:
$
\ket{\Psi_{\mathrm{SILC}}}
=
\prod_q
\prod_{m=-N_\phi/2}^{N_\phi/2}
c_{q,\sigma,m}^\dagger
\ket{0},
$
where
$
c_{q,\sigma,m}
=
\frac{1}{\sqrt{N_L}}
\sum_l
e^{iql}
c_{l,\sigma,m}$ is an electron operator with momentum $q$ along the magnetic-field direction and the Landau-level orbital index $m$. This can be straightforwardly generalized to valley-coherent states or other valley-ordered configurations (Supplementary Note 2.3).

Another important competing phase is provided by the staged FQH states (Fig.~\ref{fig2}b). These states correspond to configurations in which electrons selectively occupy only a subset of layers, thereby forming a charge-density-wave order along the magnetic-field direction. For example, a staged Laughlin state can be expressed as 
$
\Psi
\sim
\prod_{l\in 3\mathbb{Z}}
\prod_{i<j}
(z_{l,\sigma,i}-z_{l,\sigma,j})^{3}
$. Here, electrons populate only one layer among every three consecutive layers, with each occupied layer independently realizing the conventional two-dimensional Laughlin state. This can be straightforwardly extended to other filling fractions, valley-coherent configurations, and more elaborate layer-resolved ordering patterns with nonuniform electron densities across different layers (Supplementary Note 2.6). 

\paragraph{Phase Diagrams.}\noindent
Based on these candidate wavefunctions, we perform Monte Carlo to identify the ground states in the conventional three-dimensional graphite limit and in alternating and helical twisted multilayer graphene. The filling fraction per layer is defined including the valley degeneracy: 
$
\nu_L = \frac{1}{2L}\left(\frac{N_e}{N_\Phi+S}\right)
$ 
where $L$ is the number of layers, \(N_e\) is the number of electrons, \(N_\Phi\) is the number of flux quanta, and \(S\) is the shift. In this work, we focus on the fractional filling $\nu_L<1$. The parameters Eq.~\eqref{eq:Hamiltonian} are adopted from models known to reproduce the experimentally observed phase diagrams of large-angle twisted bilayer and trilayer graphene under strong magnetic fields~\cite{NatComm2017_Hunt, NanoLett2023_Kim, NatComm2025_Kim, NanoLett2026_Kim} (Supplementary Table 5). The resulting phase diagrams (Fig.~\ref{fig3}) reveal several notable features.

First, the large magnetic-field regime $B \gtrsim 100~\mathrm{T}$ provides an important consistency check for both our calculations and physical intuition, although this parameter range lies beyond experimental accessibility. In this strong-field regime, corresponding to $d/l_B \gtrsim 0.2$ with $d$ the interlayer atomic spacing and \(l_B\) the magnetic length, the influence of the twist becomes effectively negligible and all three systems exhibit nearly identical phase diagrams. This behavior is naturally understood from the dominance of the intralayer Coulomb interaction in this regime, which renders the low-energy physics effectively two-dimensional and thus insensitive to whether the neighboring layers are twisted. Consistent with this, the strong-field phase diagrams (Fig.~\ref{fig3}a--c) are dominated by staged Laughlin states and their particle-hole conjugates, which are essentially stacks of conventional two-dimensional quantum Hall states.  

The impact of the twist on the phase diagram becomes most pronounced in the experimentally accessible regime of magnetic fields, i.e., $B \lesssim 20~\mathrm{T}$. First, in the untwisted limit (Fig.~\ref{fig3}a), the phase diagram is largely dominated by metallic SILC states as expected. In the twisted systems (Fig.~\ref{fig3}b,c), the SILC phases are dramatically suppressed, accompanied by the concurrent emergence of generalized Halperin states. The generalized Halperin states are particularly robust at \(\nu_L=1/7\) and \(\nu_L=1/9\). Remarkably, they remain stable even at the comparatively large filling fraction \(\nu_L=1/2\), requiring magnetic fields of only \(B\approx3~\mathrm{T}\) for alternating twisted multilayers (\(t_\perp\approx1.78~\mathrm{meV}\)) and as low as \(B\approx1~\mathrm{T}\) for the helical twisted case (\(t_\perp\approx1.09~\mathrm{meV}\)), consistent with our physical intuition. Their valley ordering and entanglement structures are illustrated in Fig.~\ref{fig2}c--e.

Several remarks are in order. First, the overall topology, phase structure, and qualitative shape of the phase diagrams remain robust against modest variations of the model parameters (Supplementary Note 3). This robustness further strengthens our conclusions and supports the experimental observability of the generalized Halperin states. Second, the emergence of generalized Halperin states together with the suppression of metallic SILC phases in large-angle twisted systems relative to the untwisted limit is not specific to graphene. Indeed, for transition metal dichalcogenides, we obtain qualitatively similar phase diagrams (Supplementary Note 4). 

\paragraph{Anyons with Irrational Statistics.}\noindent
We next proceed to analyze their quasiparticle excitations using the infinite Chern-Simons theory developed in~\cite{ PRB2022_Ma} (Supplementary Note 5). To begin, we consider the generalized Halperin state $(3110)$ at $\nu_L=1/7$ (Fig.~\ref{fig2}d). This phase has rationally quantized Hall conductivity $\sigma_{xy}=2e^2/7h$ and shift $S=3$. Most strikingly, quasiparticle excitations of rational electric charge $-e/7$ exhibit irrational self-statistical angles,
\begin{align}
\theta=\pi\sqrt{\frac{1}{7}+\frac{2}{3\sqrt{21}}},
\end{align}
a phenomenon impossible within strictly two-dimensional topological orders. The mutual braiding statistics between quasiparticles in different layers are likewise irrational (Supplementary Note 5). Importantly, irrational anyonic excitations are not unique to this particular example, but emerge generically across different generalized Halperin states appearing in our phase diagrams, whose details are in Supplementary Note 5. 

It is natural to ask how the irrational statistics in the infinite-layer limit are modified in finite-layer systems, since experimentally realized twisted vdW multilayers necessarily contain a finite number of layers, albeit potentially as many as hundreds~\cite{Nature2024_Ji}. Remarkably, the anyon statistics converge rapidly to their infinite-layer values as the number of layers increases, with deviations $< 0.1$ rad once the system contains more than $20$ layers~(Supplementary Figure 6). Since such layer numbers are already experimentally accessible~\cite{Nature2024_Ji, Science2020_Zhao, NatComm2025_Zhang}, our results suggest a practical pathway toward observing irrational statistics and non-foliated fracton orders in real materials.    

\section{Discussion}\noindent
We demonstrated that large-angle twisted vdW multilayers under strong magnetic fields provide a realistic platform for realizing three-dimensional non-foliated FQH phases, unavailable in conventional stacks of decoupled two-dimensional quantum Hall states. Performing Monte Carlo over $862$ physically motivated Ansatz wavefunctions, we showed that the strong suppression of interlayer tunneling by the twist, together with the strong interlayer Coulomb interaction, stabilizes generalized Halperin states with interlayer quantum coherence across multiple consecutive layers. These phases emerge in experimentally accessible magnetic-field regimes where untwisted three-dimensional systems are instead dominated by metallic SILC states. We further established that these generalized Halperin states possess quasiparticle excitations with irrational braiding statistics despite carrying rational electric charges and quantized Hall responses, thereby realizing a fundamentally new form of fractionalization beyond conventional topological order. 

Several future directions follow from this work. First, since anomalous FQH states in twisted vdW materials have recently attracted significant attention~\cite{Nature2023_Cai, Nature2023_Park, PRB2023_Reddy, PRL2024_Wang, PRB2024_Ahn, NatComm2025_Chen, NatComm2026_Ahn}, it would be interesting to explore whether anomalous counterparts of generalized Halperin states can emerge in the three-dimensional limit. Second, our phase diagrams contain multiple competing phases and transitions whose nature and critical properties remain to be understood. Finally, the generalized Halperin states studied here possess infinitely many topological ground-state degeneracy in the idealized infinite-layer limit. In realistic materials this degeneracy is expected to become finite, though potentially still exponentially large, raising intriguing possibilities for applications to quantum memories and topological quantum information processing \cite{QPL2009_Wootton, RevModPhys2016_Brown, MPA2020_Pretko}. Our work suggests that twisted vdW multilayers may provide the first realistic condensed-matter platform for non-foliated topological phases and their associated topological degeneracies. 

\section{Acknowledgments}\noindent 
We thank Youngwoo Son for helpful discussions. S.~J., H.~L., G.~Y.~C., acknowledge the financial supports by Samsung Science and Technology Foundation under Project Number SSTF-BA2401-03, the NRF of Korea (Grants No. RS-2026-25479545, RS-2024-00410027, RS-2023-NR119931, RS-2024-00444725, RS-2023-00256050, RS-2025-25453111, RS-2025-08542968) funded by the Korean Government (MSIT), the Air Force Office of Scientific Research under Award No. FA23862514026, and Institute of Basic Science under project code IBS-R014-D1. Y.~K. was supported by NRF (Grant No. RS-2025-00557717, RS-2023-00269616, No. RS-2024-00444725, RS-2025-02317602, RS-2025-02317602) funded by MSIT. Y.~K. also acknowledges the partner group program of the Max Planck Society.   

\section{Methods} 
\paragraph{Details of the model}\noindent
The model Eq.~\eqref{eq:Hamiltonian} contains two degenerate Landau levels per layer, representing the \(K\) and \(K'\) valley degrees of freedom. We consider twist angles in the range \(0 \leq \theta<30^\circ\). The Coulomb interaction is given by  
\begin{align}
    H_{\text{Coulomb}} = \frac{1}{2} &\sum_{l,l',\sigma,\sigma'} \int d\mathbf{r} d\mathbf{r}' \; \hat{\psi}^\dagger_{l,\sigma}(\mathbf{r}) \hat{\psi}^\dagger_{l',\sigma'}(\mathbf{r}') \nonumber \\ 
        &~~~~~~~~~~~~V_{ll'\sigma\sigma'}(\mathbf{r},\mathbf{r}') \hat{\psi}_{l',\sigma'}(\mathbf{r}') \hat{\psi}_{l,\sigma}(\mathbf{r}), \nonumber  
\end{align}
where \(\hat{\psi}^\dagger_{l,\sigma}(\mathbf{r})\) (\(\hat{\psi}_{l,\sigma}(\mathbf{r})\)) is the electron creation (annihilation) operator for an electron in layer \(l \in \mathbb{Z}\) and valley \(\sigma \in \{K, K'\}\) at the two-dimensional coordinate \(\mathbf{r}\) and  
\begin{equation*}
V_{ll'\sigma\sigma'}(\mathbf{r},\mathbf{r}') = \frac{e^2}{\epsilon l_B} \frac{1}{\sqrt{\left(|\mathbf{r}-\mathbf{r}'|/l_B\right)^2+\left(|l-l'|d/l_B\right)}}.
\end{equation*}
The tunneling Hamiltonian \(H_{\mathrm{Tunnel}}\) retains only interlayer intravalley tunneling as the leading contribution. Although interlayer intervalley tunneling is allowed in principle, it is expected to be strongly suppressed by the large momentum mismatch between the \(K\) and \(K'\) valleys and is neglected. Finally, the short-ranged interaction is 
\begin{align}
	H_V = \sum_{l} \int d\mathbf{r} \ \Big(  V_1~ \hat{\psi}^\dagger_{l,K}(\mathbf{r}) \hat{\psi}^\dagger_{l,-K}(\mathbf{r})\hat{\psi}_{l,-K}(\mathbf{r})\hat{\psi}_{l,K}(\mathbf{r})\nonumber \\
	+ \sum_\sigma \Big[V_2 ~\hat{\psi}^\dagger_{l,\sigma}(\mathbf{r}) \hat{\psi}^\dagger_{l+1,\sigma}(\mathbf{r})\hat{\psi}_{l+1,\sigma}(\mathbf{r})\hat{\psi}_{l,\sigma}(\mathbf{r})\nonumber \\
	+ V_3~ \hat{\psi}^\dagger_{l,\sigma}(\mathbf{r}) \hat{\psi}^\dagger_{l+1,-\sigma}(\mathbf{r})\hat{\psi}_{l+1,-\sigma}(\mathbf{r})\hat{\psi}_{l,\sigma}(\mathbf{r})\Big]\Big). \nonumber
\end{align}
where \(V_1\), \(V_2\), and \(V_3\) are chosen to reproduce the experimentally observed phase diagrams~\cite{NanoLett2023_Kim, NatComm2025_Kim, NanoLett2026_Kim}. 

\paragraph{Construction of Ansatz wavefunctions.}\noindent
Our variational space consists of \(696\) liquid states and \(166\) crystalline states. Starting from a set of elementary liquid and crystalline ``seed'' wavefunctions, which cannot be generated from one another through products, particle-hole conjugation, or staging, we systematically generate the remaining candidate states by combining these operations with possible valley-ordering patterns. Here, a staged state~\cite{PRB1988_MacDonald, PRB2009_Burnell} denotes a state whose charge density is modulated along the stacking direction. The Ansatz wavefunctions is built from the \(43\) liquid and \(123\) crystalline seed wavefunctions of distinct flux-attachment patterns and filling fractions. The detailed forms of liquid and crystalline seed states are described in Supplementary Note 2. For crystalline states, both the many-body wavefunction and the crystal geometry must be specified. We determine the optimal crystal geometries by numerically solving the spherical Thomson problem~\cite{PRB1999_PerezGarrido, PRL2013_Archer} using the basin-hopping algorithm~\cite{PRB2018_Faugno, PRB2006_Wales} (Supplementary Note 2.2).

\paragraph{Monte Carlo calculation of energy.}\noindent
Monte Carlo calculations are performed to evaluate the Coulomb energies of the candidate liquid and crystalline Ansatz wavefunctions, i.e., generalized Halperin, staging states, and composite-fermion crystals. The simulations employ a fundamental cell consisting of \(N_L\) layers with periodic boundary conditions along the stacking direction. Each layer contains \(48\) electrons, and for valley-coherent states the electrons are assumed to be equally distributed between the \(K\) and \(K'\) valleys. For each choice of \(d/l_B\) and \(N_L\), the Coulomb energy per particle, \(E_{\mathrm{Coulomb}}/N_{\mathrm{tot}}\), is evaluated using Metropolis--Hastings sampling~\cite{NatComm2025_Kim} over \(10^6\) particle configurations. The acceptance probability is determined by the ratio of the probability amplitudes of the candidate wavefunction. For each simulation, the algorithm is optimized using prerun estimates of the acceptance rate and integrated autocorrelation time. For each sampled configuration, the Coulomb energy is evaluated using the approximation scheme described in Supplementary Note 6, with a relative cutoff parameter \(M_{\mathrm{eff}}\). The value of \(M_{\mathrm{eff}}\) is calibrated using benchmark calculations for decoupled stacks of \(1/3\) Laughlin states to ensure convergence. The Coulomb energy is computed for a range of \(N_L\), and the thermodynamic-limit value is obtained by least-squares extrapolation of \(E_{\mathrm{Coulomb}}/N_{\mathrm{tot}}\) as a function of \(1/N_L\) to \(N_L\rightarrow\infty\). Further details of the Monte Carlo procedure are provided in Supplementary Note 6. The calculation of energy for the metallic spontaneous interlayer coherent states is in Supplementary Note 2.3.

Several notes are in order. First, the energies of product and particle-hole-conjugate states can be obtained directly from those of the corresponding seed states using the relations derived in Supplementary Note 2.4 and Supplementary Note 2.5, respectively. The energies of staged states are computed by adding the corresponding staging energies, which account for the energy cost associated with the nonuniform charge distribution along the stacking direction. We consider four distinct staging patterns, and their staging energies are evaluated as described in Supplementary Note 2.6. Second, as shown in Supplementary Note 6, whenever a state is constructed as a double copy of the generalized Halperin state—such as those appearing in the phase diagrams of Fig.~\ref{fig3}—each generalized Halperin state exhibits a valley polarization, which means that one liquid occupies valley \(K\) and the other occupies \(K'\).

\bibliography{refs.bib}

\end{document}


\title{Supplementary Information for ``Three-Dimensional Non-Foliated Fractional Quantum Hall Phases with Irrational Anyons in Twisted van der Waals Multilayers''}

\author{Seyoung Jin}
\affiliation{Department of Physics, Pohang University of Science and Technology, Pohang, 37673, Republic of Korea}

\author{Hyeonseo Lim}
\affiliation{Department of Physics, Pohang University of Science and Technology, Pohang, 37673, Republic of Korea}

\author{Youngwook Kim}
\thanks{y.kim@dgist.ac.kr}
\affiliation{Department of Physics and Chemistry, Daegu Gyeongbuk Institute of Science and Technology (DGIST), Daegu 42988, Republic of Korea}

\author{Gil Young Cho}
\thanks{gilyoungcho@kaist.ac.kr}
\affiliation{Department of Physics, Korea Advanced Institute of Science and Technology, Daejeon 34141, Republic of Korea}
\affiliation{Center for Artificial Low Dimensional Electronic Systems, Institute for Basic Science, Pohang 37673, Korea}

\begingroup
\setlength{\parskip}{0pt} 
\maketitle
\tableofcontents
\endgroup

\supnotesection{Geometry of the System}

We consider a stack of infinite layers which are equally spaced so that the interlayer distance between any two adjacent layers is \(d\). In particular, we construct the `fundamental cell' consisting of \(N_L\) layers and replicate it under periodic boundary condition along the stacking axis. A uniform external magnetic field is applied in the direction perpendicular to the layers.

Instead of a flat surface, we map each layer onto a sphere. The Haldane sphere encloses a magnetic monopole, which generates the magnetic field \(\mathbf{B}=B\hat{\mathbf{r}}\). The coordinates of a particle on the sphere is given in terms of the spinor variables \(u = \cos(\vartheta/2)\exp(i\varphi/2)\), and \(v = \sin(\vartheta/2)\exp(-i\varphi/2)\) with the polar angle \(\vartheta\) and the azimuthal angle \(\varphi\) in the spherical coordinate system. Moreover, for any positions \(\mathbf{r}\) and \(\mathbf{r}'\) on the sphere, we define the chord distance \cite{Paper_Morf} \(|\mathbf{r}_i-\mathbf{r}_j| = 2R|uv'-vu'|\), where \(R = l_B\sqrt{N_\Phi/2}\) is the radius of Haldane sphere, \(l_B = \sqrt{\frac{\hbar c}{eB}}\) is the magnetic length, and \(N_\Phi\) is the number of flux quanta of our interest. 

\supnotesection{Candidate Wavefunctions}
Exponential factors in the wavefunctions below are omitted.

\paragraph{Local Valley Polarization} If the state resides in a single valley on each layer but not necessarily in the same valley throughout the system, it is referred as locally valley-polarized.

\subsection{Generalized Halperin Liquids}
\paragraph{With Local Valley Polarization} We consider the generalized Halperin liquids proposed in \cite{PRB1990_Qiu, PRL2000_Naud, NPB2001_Sondhi, PRB2009_Burnell}, which occupies a single valley on each layer. Letting \(N_L\) be number of layers in the `fundamental cell', each state labeled by \((mnop)\) is characterized by a \(N_L\times N_L\) matrix \(K\), and two vectors of \(N_L\) elements, \(\mathbf{t}\), and \(\mathbf{s}\):
\begin{equation} \label{liquidkmatrix}
    K = 
    \begin{pmatrix}
        m & n & o & p & 0 & 0 & 0 & \cdots & p & o & n \\ 
        n & m & n & o & p & 0 & 0 & \cdots & 0 & p & o \\
        o & n & m & n & o & p & 0 & \cdots & 0 & 0 & p \\
        p & o & n & m & n & o & p & \cdots & 0 & 0 & 0 \\
        \vdots & \vdots & \vdots & \vdots & \vdots & \vdots & \vdots &  & \vdots & \vdots & \vdots \\
        p & 0 & 0 & 0 & 0 & 0 & 0 & \cdots & m & n & o \\
        o & p & 0 & 0 & 0 & 0 & 0 & \cdots & n & m & n \\
        n & o & p & 0 & 0 & 0 & 0 & \cdots & o & n & m \\
    \end{pmatrix},\quad
    \mathbf{t} = 
    \begin{pmatrix}
        1 \\
        1 \\
        1 \\
        1 \\
        \vdots \\
        1 \\
        1 \\
        1 \\        
    \end{pmatrix},\
    \mathbf{s} = 
    \begin{pmatrix}
        m/2 \\
        m/2 \\
        m/2 \\
        m/2 \\
        \vdots \\
        m/2 \\
        m/2 \\
        m/2 \\        
    \end{pmatrix},
\end{equation}
i.e., for each \(1\leq i,j\leq N_L\), 
\begin{equation}
    K_{ij} = m\delta_{i,j} + n\left(\delta_{i,j-1}+\delta_{i,j+1}\right) + o\left(\delta_{i,j-2}+\delta_{i,j+2}\right) + p\left(\delta_{i,j-3}+\delta_{i,j+3}\right),\quad t_i = 1,\quad s_i=\frac{m}{2}.
\end{equation}
In order for the electrons to have Fermi statistics, \(m\in2\mathbb{Z}+1\) and \(n,o,p\in\mathbb{Z}\) \cite{Book_Wen}. Neglecting the valley indices, the wavefunction is provided as
\begin{equation}
    \Psi(\{z\}) = \prod_{l}^{N_L} \Biggl(\Biggl[\prod_{i<j} (z_{l,i}-z_{l,j})^{K_{ll}}\Biggr]\cdot\Biggl[\prod_{l'>l}^{N_L} \prod_{i,j} (z_{l,i}-z_{l',j})^{K_{ll'}}\Biggr]\Biggr).
\end{equation}
That is, letting \(l+N_L\equiv l\), the wavefunction of the \((mnop)\) state is
\begin{equation}
    \Psi(\{z\}) 
    = \prod_{l}^{N_L} \Biggl(\Biggl[\prod_{i<j} (z_{l,i}-z_{l,j})^{m}\Biggr]\cdot\Biggl[\prod_{i,j} (z_{l,i}-z_{l+1,j})^{n}\Biggr]\cdot\Biggl[\prod_{i,j} (z_{l,i}-z_{l+2,j})^{o}\Biggr]\cdot\Biggl[\prod_{i,j} (z_{l,i}-z_{l+3,j})^{p}\Biggr]\Biggr).
\end{equation}
Moreover, from the above characterization, we obtain the average filling fraction per layer \(\bar{\nu}\), and the shift \(S\):
\begin{equation}
    \bar{\nu} = \frac{1}{N_L}\transpose{\mathbf{t}}K^{-1}\mathbf{t} = \frac{1}{m+2n+2o+2p},\quad S = \frac{2\transpose{\mathbf{t}}K^{-1}\mathbf{s}}{\transpose{\mathbf{t}}K^{-1}\mathbf{t}}= m.
\end{equation}
Letting \(\bar{N}\) be the average number of particles per layer, \(N_\Phi = \bar{N}/\bar{\nu}-S\) in the spherical geometry. We restrict \(m \geq n \geq o \geq p\), since interlayer coherence are expected to decrease with increasing layer separation. Locally valley-polarized generalized Halperin states we consider are listed in \ref{TableS1}. 

\begin{table}[H]
\begin{center}
\begin{minipage}[t]{0.148\textwidth}
\begin{center}
\begin{tabularx}{\textwidth}{| c | c | c |}
    \hline
    \(\bar{\nu}\) & \((mnop)\) & \(S\) \\
    \hline \hline
    1                    & (1000) & 1 \\ \hline
    \multirow{2}{*}{1/3} & (3000) & 3 \\
                         & (1100) & 1 \\
    \hline
    \multirow{3}{*}{1/5} & (5000) & 5 \\
                         & (3100) & 3 \\
                         & (1110) & 1 \\
    \hline
\end{tabularx}
\end{center}
\end{minipage}
\hspace{-4pt}
\begin{minipage}[t]{0.148\textwidth}
\begin{center}
\begin{tabularx}{\textwidth}{| c | c | c |}
    \hline
    \(\bar{\nu}\) & \((mnop)\) & \(S\) \\
    \hline \hline
    \multirow{5}{*}{1/7} & (7000) & 7 \\
                         & (5100) & 5 \\
                         & (3200) & 3 \\
                         & (3110) & 3 \\
                         & (1111) & 1 \\
    \hline
    1/9 & (9000) & 9 \\
    \hline
\end{tabularx}
\end{center}
\end{minipage}
\hspace{-4pt}
\begin{minipage}[t]{0.148\textwidth}
\begin{center}
\begin{tabularx}{\textwidth}{| c | c | c |}
    \hline
    \(\bar{\nu}\) & \((mnop)\) & \(S\) \\
    \hline \hline
    \multirow{6}{*}{1/9} & (7100) & 7 \\
                         & (5200) & 5 \\
                         & (5110) & 5 \\
                         & (3300) & 3 \\
                         & (3210) & 3 \\
                         & (3111) & 3 \\
    \hline
\end{tabularx}
\end{center}
\end{minipage}
\caption{List of locally valley-polarized generalized Halperin liquids \((mnop)\), with filling fraction per layer \(\bar{\nu}\) and their shift \(S\).}
\label{TableS1}
\end{center}
\end{table}

\paragraph{Without Local Valley Polarization} Similarly, we consider the state with \(m\) intralayer-intravalley zeros, \(n\) intralayer-intervalley zeros, \(o\) interlayer-intravalley zeros, and \(p\) interlayer-intervalley zeros. Denoted by \((mn/op)\), such liquid state is characterized by the new matrix \(K\) with the same \(\mathbf{t}\) and \(\mathbf{s}\) in Eq.~\eqref{liquidkmatrix}:
\begin{equation}
    K = 
    \begin{pmatrix}
        K_{intra} & K_{inter} & \mathbf{0}   & \cdots & \mathbf{0}   & K_{inter} \\
        K_{inter} & K_{intra} & K_{inter} & \cdots & \mathbf{0}   & \mathbf{0}   \\
        \mathbf{0}   & K_{inter} & K_{intra} & \cdots & \mathbf{0}   & \mathbf{0}   \\
        \vdots    & \vdots    & \vdots    & \ddots & \vdots    & \vdots    \\
        \mathbf{0}   & \mathbf{0}   & \mathbf{0}   & \cdots & K_{intra} & K_{inter} \\
        \mathbf{0}   & \mathbf{0}   & \mathbf{0}   & \cdots & K_{inter} & K_{intra} \\
    \end{pmatrix}, \quad
    K_{intra} = 
    \begin{pmatrix}
        m & n \\
        n & m \\
    \end{pmatrix},\quad
    K_{inter} = 
    \begin{pmatrix}
        o & p \\
        p & o \\
    \end{pmatrix},
\end{equation}
i.e., for each \(1\leq i,j\leq N_L\) and each \(k,l\in\{1,2\}\),
\begin{equation}
    K_{2i+k,2j+l} = m\delta_{i,j}\delta_{k,l} + n\delta_{i,j}\left(1-\delta_{k,l}\right) + o\left(\delta_{i,j-1}+\delta_{i,j+1}\right)\delta_{k,l} + p\left(\delta_{i,j-1}+\delta_{i,j+1}\right)\left(1-\delta_{k,l}\right).
\end{equation}
That is, the wavefunction of the \((mn/op)\) state is
\begin{align}
    \Psi(\{z\}) 
    = \prod_{l=1}^{N_L} \Biggl(&\Biggl[\prod_\sigma \prod_{i<j}(z_{l,\sigma,i}-z_{l,\sigma,j})^{m}\Biggr]\cdot\Biggl[\prod_{i,j} (z_{l,K,i}-z_{l,K',j})^{n}\Biggr]\nonumber \\&\cdot\Biggl[\prod_\sigma \prod_{i,j} (z_{l,\sigma,i}-z_{l+1,\sigma,j})^{o}\Biggr]\cdot\Biggl[\prod_\sigma \prod_{i,j} (z_{l,\sigma,i}-z_{l+1,-\sigma,j})^{p}\Biggr]\Biggr).
\end{align}
where \(K'=-K\). Moreover, from the above characterization, we obtain the average filling fraction per layer per valley \(\nu_L\), and the shift \(S\):
\begin{equation}
    \nu_L = \frac{1}{2 N_L}\transpose{\mathbf{t}}K^{-1}\mathbf{t} = \frac{1}{m+n+2o+2p},\text{ and } S = \frac{2\transpose{\mathbf{t}}K^{-1}\mathbf{s}}{\transpose{\mathbf{t}}K^{-1}\mathbf{t}}= m.
\end{equation}
Letting \(N\) be the average number of particles per layer per valley, \(N_\Phi = N/\nu_L-S\) in the spherical geometry. Since the interlayer interaction can be stronger than the intralayer-intervalley interaction, we set \(m \geq n, o, p\). \((mn/op)\) states of our interest are provided in \ref{TableS2}.

\begin{table}[H]
\begin{center}
\begin{minipage}{0.214\textwidth}
\begin{center}
\begin{tabularx}{\textwidth}{| c | c | c |}
    \hline
    \(\nu_L\) & \((mn/op)\) & \(S\) \\
    \hline \hline
    1/2                  & (11/00) & 1  \\ 
    \hline
    \multirow{2}{*}{1/4} & (31/00) & 3  \\ 
                         & (11/10) (11/01) & 1 \\ 
    \hline
    \multirow{2}{*}{1/5} & (32/00) & 3 \\ 
                         & (10/11) & 1 \\ 
    \hline
    \multirow{4}{*}{1/6} & (51/00) & 5 \\ 
                         & (33/00) & 3 \\ 
                         & (31/10) (31/01)& 3 \\ 
                         & (11/11) & 1 \\ 
    \hline
\end{tabularx}
\end{center}
\end{minipage}
\hspace{-4pt}
\begin{minipage}{0.214\textwidth}
\begin{center}
\begin{tabularx}{\textwidth}{| c | c | c |}
    \hline
    \(\nu_L\) & \((mn/op)\) & \(S\) \\
    \hline \hline
    \multirow{3}{*}{1/7} & (52/00) & 5 \\ 
                         & (32/10) (32/01) & 3  \\ 
                         & (30/11) & 3          \\ 
    \hline
    \multirow{6}{*}{1/8} & (71/00) & 7  \\ 
                         & (53/00) & 5  \\ 
                         & (51/10) (51/01) & 5\\ 
                         & (33/10) (33/01) & 3  \\ 
                         & (31/20) (31/02) & 3  \\
                         & (31/11) & 3  \\
    \hline
\end{tabularx}
\end{center}
\end{minipage}
\hspace{-4pt}
\begin{minipage}{0.214\textwidth}
\begin{center}
\begin{tabularx}{\textwidth}{| c | c | c |}
    \hline
    \(\nu_L\) & \((mn/op)\) & \(S\) \\
    \hline \hline
    \multirow{7}{*}{1/9} & (72/00) & 7 \\ 
                         & (54/00) & 5 \\ 
                         & (52/10) (52/01) & 5 \\ 
                         & (50/11) & 5  \\ 
                         & (32/20) (32/02)& 3  \\ 
                         & (32/11) & 3 \\ 
                         & (30/21) (30/12) & 3  \\ 
    \hline
\end{tabularx}
\vspace{33pt}
\end{center}
\end{minipage}
\caption{List of generalized Halperin liquids without local valley polarization \((mn/op)\), along with their filling fraction per layer per valley \(\nu_L\) and shift \(S\). States which can be constructed from the states in \ref{TableS1} are not listed.}
\label{TableS2}
\end{center}
\end{table}

\subsection{Generalized Composite Fermion Crystals}
We define crystal sites in uppercase. In particular, the coordinates of a lattice site is given in terms of the spinor variables \(U = \cos(\varTheta/2)\exp(i\varPhi/2)\), and \(V = \sin(\varTheta/2)\exp(-i\varPhi/2)\) with the polar angle \(\varTheta\) and the azimuthal angle \(\varPhi\) in the spherical geometry. To determine the crystal configuration, we solve the spherical Thomson problem \cite{PRB2018_Faugno, PRL2013_Archer}, which is to find the lowest energy configuration of given number of classical point charges confined on the sphere \cite{PRB1999_PerezGarrido, PRB2018_Faugno, PRL2013_Archer}. The distance between a pair of charges at spinor coordinates \((u_i, v_i)\) on layer \(l_i\) and \((u_j, v_j)\) on layer \(l_j\) is defined to be
\begin{equation}
    r_{i,j} = \sqrt{|u_iv_j-v_iu_j|^2 + \min((l_i-l_j)\bmod N_L,(l_j-l_i)\bmod N_L)^2(d/2R)^2},
\end{equation}
where \(N_L\) is the number of layers in the periodic cell, and \(R\) is the radius of the Haldane sphere.

\paragraph{Numerical Solution of the Spherical Thomson Problem} To determine optimal crystal configurations, spherical Thomson problems are numerically solved in the basin-hopping method. In particular, Optim.jl \cite{Optim}, Linesearches.jl, and Basinhopping.jl packages \cite{Julia} are used to implement the solver. In the basin-hopping algorithm \cite{JPhysChemA1997_Wales}, \(5\times 10^4\) local optimizations are performed. As an optimizer for the basin-hopping method \cite{JChemPhys2004_Trygubenko, PRB2018_Faugno}, we use limited-memory Broyden-Fletcher-Goldfarb-Shanno (L-BFGS) algorithm \cite{MP1989_Liu, Book_Nocedal}, along with backtracking \cite{Book_Nocedal} as its line search algorithm. The number of L-BFGS iterations is set to be at most \(10^3\).

\paragraph{With Local Valley Polarization} A single-component composite fermion crystal (CFC) wavefunction is given by \cite{PRB2018_Faugno, PRL2013_Archer}
\begin{equation}
    \Psi_{\tilde{\nu}}^{\mathrm{Crys}(2p)}(\{z_i\}) = \det(U_i^*u_j+V_i^*v_j)^{N_\phi^*}\prod_{i<j}(u_iv_j-v_iu_j)^{2p}
\end{equation}
where \(N_\phi^* = \tilde{N_\phi} - 2p(N-1)\). Here, \(\lim\limits_{n \to \infty} \bar{N} / \tilde{N_\phi} = \tilde{\nu}\) where \(\bar{N}\) is the number of particles (per layer). Generalizing the above, the wavefunction of the crystal \(\textrm{Crys}(2p,m,n,o)\) with filling fraction per layer \(\bar{\nu}\) is
\begin{equation}
    \Psi(\{z\}) = \prod_{i=1}^{N_L} \Biggl(\Psi_{\frac{\bar{\nu}}{1-2(m+n+o)\bar{\nu}}}^{\textrm{Crys}(2p)}(\{z_{l,i}\}) \cdot \Biggl[\prod_{i,j} (z_{l,i}-z_{l+1,j})^m\Biggr] \cdot \Biggl[\prod_{i,j} (z_{l,i}-z_{l+2,j})^n\Biggr] \cdot \Biggl[\prod_{i,j} (z_{l,i}-z_{l+3,j})^o\Biggr] \Biggr)
\end{equation}
with valley indices omitted, while \(\tilde{N_\phi} = N_\phi - 2(m+n+o)\bar{N}\) for \(\tilde{\nu}=\bar{\nu}/(1-2(m+n+o)\bar{\nu})\). As done for the liquid states, we restrict \(2p \geq m \geq n \geq o\) to be physical. Moreover, we construct a one-to-one mapping between \((mnop)\) and \(\textrm{Crys}(2p, m, n, o)\) with the same \(\bar{\nu}\), and assume that the number of flux quanta of each liquid state and the corresponding crystal state are the same. Then, there is no contradiction to set \(N_\phi = \bar{N}/\bar{\nu} - 2p - 1\). \ref{TableS3} is the list of \(\mathrm{Crys}(2p,m,n,o)\) we consider as candidates.

\begin{table}[H]
\begin{center}
\begin{minipage}{0.206\textwidth}
\begin{center}
\begin{tabularx}{\textwidth}{| c | c |}
    \hline
    \(\bar{\nu}\) & \(\textrm{Crys}(2p,m,n,o)\) \\
    \hline \hline
    1   & \multirow{3}{*}{\(\textrm{Crys}(0,0,0,0)\)} \\
    2/3 &                                 \\
    1/2 &                                 \\
    \hline
    2/5 & \\
    1/3 & \(\textrm{Crys}(0,0,0,0)\) \\
    2/7 & \(\textrm{Crys}(2,0,0,0)\) \\
    1/4 & \\
    \hline
    \multirow{4}{*}{\minitab[c]{2/9\\1/5\\1/6}}& \(\textrm{Crys}(0,0,0,0)\) \\
    & \(\textrm{Crys}(2,0,0,0)\) \\
    & \(\textrm{Crys}(2,1,0,0)\) \\
    & \(\textrm{Crys}(4,0,0,0)\) \\
    \hline
\end{tabularx}
\end{center}
\end{minipage}
\hspace{-4pt}
\begin{minipage}{0.206\textwidth}
\begin{center}
\begin{tabularx}{\textwidth}{| c | c |}
    \hline
    \(\bar{\nu}\) & \(\textrm{Crys}(2p,m,n,o)\) \\
    \hline \hline
        & \(\textrm{Crys}(0,0,0,0)\) \\
        & \(\textrm{Crys}(2,0,0,0)\) \\
        & \(\textrm{Crys}(2,1,0,0)\) \\
    1/7 & \(\textrm{Crys}(2,1,1,0)\) \\
    1/8 & \(\textrm{Crys}(2,2,0,0)\) \\
        & \(\textrm{Crys}(4,0,0,0)\) \\
        & \(\textrm{Crys}(4,1,0,0)\) \\
        & \(\textrm{Crys}(6,0,0,0)\) \\
    \hline
    \multirow{3}{*}{\minitab[c]{1/9}} & \(\textrm{Crys}(0,0,0,0)\) \\
        & \(\textrm{Crys}(2,0,0,0)\) \\
        & \(\textrm{Crys}(2,1,0,0)\) \\
\hline
\end{tabularx}
\end{center}
\end{minipage}
\hspace{-4pt}
\begin{minipage}{0.206\textwidth}
\begin{center}
\begin{tabularx}{\textwidth}{| c | c |}
    \hline
    \(\bar{\nu}\) & \(\textrm{Crys}(2p,m,n,o)\) \\
    \hline \hline
    \multirow{11}{*}{\minitab[c]{1/9}} & \(\textrm{Crys}(2,1,1,0)\) \\
        & \(\textrm{Crys}(2,1,1,1)\) \\
        & \(\textrm{Crys}(2,2,0,0)\) \\
        & \(\textrm{Crys}(2,2,1,0)\) \\
        & \(\textrm{Crys}(4,0,0,0)\) \\
        & \(\textrm{Crys}(4,1,0,0)\) \\
        & \(\textrm{Crys}(4,1,1,0)\) \\
        & \(\textrm{Crys}(4,2,0,0)\) \\
        & \(\textrm{Crys}(6,0,0,0)\) \\
        & \(\textrm{Crys}(6,1,0,0)\) \\
        & \(\textrm{Crys}(8,0,0,0)\) \\        
    \hline
\end{tabularx}
\end{center}
\end{minipage}
\end{center}
\caption{List of locally valley-polarized generalized composite fermion crystals Crys\((2p,m,n,o)\) with filling fraction per layer \(\bar{\nu}\).}
\label{TableS3}
\end{table}

\paragraph{Without Local Valley Polarization} Similarly, we come up with the crystal \(\textrm{Crys}(2p,m/n,o)\) with filling fraction per layer per valley \(\nu_L\), whose wavefunction is
\begin{align}
    \Psi(\{z\}) = \prod_{l=1}^{N_L} \Biggl(&\Biggl[\prod_\sigma \Psi_{\frac{\nu_L}{1-m-2(n+o)\nu_L}}^{\textrm{Crys}(2p)}(\{z_{l,\sigma,i}\})\Biggr] \cdot \Biggl[\prod_{i,j} (z_{l,K,i}-z_{l,K',j})^m\Biggr] \nonumber \\&
    \cdot \Biggl[\prod_\sigma \prod_{i,j} (z_{l,\sigma,i}-z_{l+1,\sigma,j})^n\Biggr] \cdot \Biggl[\prod_\sigma \prod_{i,j} (z_{l,\sigma,i}-z_{l,-\sigma,j})^o\Biggr] \Biggr).
\end{align}
Note that \(\tilde{N_\phi} = N_\phi - mN - 2(n+o)N\) for \(\tilde{\nu}=\nu_L/(1-m\nu_L-2(n+o)\nu_L)\), where \(N\) is the number of particles per layer per valley. As done for the liquids, we force \(2p \geq m, n, o\) to be physical. Moreover, we construct a one-to-one mapping from the liquid states \((m,n/o,p)\) and the crystal states \(\textrm{Crys}(2p, m/ n, o)\) with \(\nu_L\), and assume that the number of flux quanta of each liquid state and the corresponding crystal state is the same. Again, there is no contradiction to set \(N_\phi = N/\nu_L - 2p - 1\). We consider the generalized composite fermion crystals without local valley polarization in \ref{TableS4}.

\begin{table}[htbp]
\begin{center}
\begin{minipage}{0.19\textwidth}
\begin{center}
\begin{tabularx}{\textwidth}{| c | c |}
    \hline
    \(\nu_L\) & \(\textrm{Crys}(2p,m/n,o)\) \\
    \hline \hline
    1/2 & \(\textrm{Crys}(0,0/0,0)\) \\
    \hline
    \multirow{2}{*}{1/3} & \(\textrm{Crys}(0,0/0,0)\) \\
                         & \(\textrm{Crys}(2,0/0,0)\) \\
    \hline
        & \(\textrm{Crys}(0,0/0,0)\) \\
    1/4 & \(\textrm{Crys}(2,0/0,0)\) \\
        & \(\textrm{Crys}(2,1/0,0)\) \\
    \hline
    \multirow{6}{*}{1/5}    & \(\textrm{Crys}(0,0/0,0)\) \\
                            & \(\textrm{Crys}(2,0/0,0)\) \\
                            & \(\textrm{Crys}(2,0/1,0)\) \\
                            & \(\textrm{Crys}(2,1/0,0)\) \\
                            & \(\textrm{Crys}(2,2/0,0)\) \\
                            & \(\textrm{Crys}(4,0/0,0)\) \\
    \hline
    \multirow{8}{*}{1/6}    & \(\textrm{Crys}(0,0/0,0)\) \\
                            & \(\textrm{Crys}(2,0/0,0)\) \\
                            & \(\textrm{Crys}(2,0/1,0)\) \\
                            & \(\textrm{Crys}(2,1/0,0)\) \\
                            & \(\textrm{Crys}(2,1/1,0)\) \\
                            & \(\textrm{Crys}(2,2/0,0)\) \\
                            & \(\textrm{Crys}(4,0/0,0)\) \\
                            & \(\textrm{Crys}(4,1/0,0)\) \\
    \hline
    \multirow{3}{*}{1/7}   & \(\textrm{Crys}(0,0/0,0)\) \\
                            & \(\textrm{Crys}(2,0/0,0)\) \\
                            & \(\textrm{Crys}(2,0/1,0)\) \\
    \hline
\end{tabularx}
\end{center}
\end{minipage}
\hspace{-4pt}
\begin{minipage}{0.19\textwidth}
\begin{center}
\begin{tabularx}{\textwidth}{| c | c |}
    \hline
    \(\nu_L\) & \(\textrm{Crys}(2p,m/n,o)\) \\
    \hline \hline
     \multirow{10}{*}{1/7}   & \(\textrm{Crys}(2,0/2,0)\) \\
                            & \(\textrm{Crys}(2,1/0,0)\) \\
                            & \(\textrm{Crys}(2,1/1,0)\) \\
                            & \(\textrm{Crys}(2,2/0,0)\) \\
                            & \(\textrm{Crys}(2,2/1,0)\) \\
                            & \(\textrm{Crys}(4,0/0,0)\) \\
                            & \(\textrm{Crys}(4,0/1,0)\) \\
                            & \(\textrm{Crys}(4,1/0,0)\) \\
                            & \(\textrm{Crys}(4,2/0,0)\) \\
                            & \(\textrm{Crys}(6,0/0,0)\) \\
    \hline
    \multirow{13}{*}{1/8}    & \(\textrm{Crys}(0,0/0,0)\) \\
                             & \(\textrm{Crys}(2,0/0,0)\) \\
                             & \(\textrm{Crys}(2,0/1,0)\) \\
                             & \(\textrm{Crys}(2,0/2,0)\) \\
                             & \(\textrm{Crys}(2,1/0,0)\) \\
                             & \(\textrm{Crys}(2,1/1,0)\) \\
                             & \(\textrm{Crys}(2,1/1,1)\) \\
                             & \(\textrm{Crys}(2,2/0,0)\) \\
                             & \(\textrm{Crys}(2,2/1,0)\) \\
                             & \(\textrm{Crys}(4,0/0,0)\) \\
                             & \(\textrm{Crys}(4,0/1,0)\) \\
                             & \(\textrm{Crys}(4,1/0,0)\) \\
                             & \(\textrm{Crys}(4,1/1,0)\) \\
                                 \hline
\end{tabularx}
\end{center}
\end{minipage}
\hspace{-4pt}
\begin{minipage}{0.19\textwidth}
\begin{center}
\begin{tabularx}{\textwidth}{| c | c |}
    \hline
    \(\nu_L\) & \(\textrm{Crys}(2p,m/n,o)\) \\
    \hline \hline
     \multirow{3}{*}{1/8}   & \(\textrm{Crys}(4,2/0,0)\) \\
                             & \(\textrm{Crys}(6,0/0,0)\) \\
                             & \(\textrm{Crys}(6,1/0,0)\) \\
    \hline
    \multirow{20}{*}{1/9}    & \(\textrm{Crys}(0,0/0,0)\) \\
                             & \(\textrm{Crys}(2,0/0,0)\) \\
                             & \(\textrm{Crys}(2,0/1,0)\) \\
                             & \(\textrm{Crys}(2,0/2,0)\) \\
                             & \(\textrm{Crys}(2,1/0,0)\) \\
                             & \(\textrm{Crys}(2,1/1,0)\) \\
                             & \(\textrm{Crys}(2,1/1,1)\) \\
                             & \(\textrm{Crys}(2,2/0,0)\) \\
                             & \(\textrm{Crys}(2,2/1,0)\) \\
                             & \(\textrm{Crys}(2,2/1,1)\) \\
                             & \(\textrm{Crys}(2,2/2,0)\) \\
                             & \(\textrm{Crys}(4,0/0,0)\) \\
                             & \(\textrm{Crys}(4,0/1,0)\) \\
                             & \(\textrm{Crys}(4,1/0,0)\) \\
                             & \(\textrm{Crys}(4,1/1,0)\) \\
                             & \(\textrm{Crys}(4,2/0,0)\) \\
                             & \(\textrm{Crys}(4,2/1,0)\) \\
                             & \(\textrm{Crys}(6,0/0,0)\) \\
                             & \(\textrm{Crys}(6,1/0,0)\) \\ 
                             & \(\textrm{Crys}(8,0/0,0)\) \\ 
    \hline
\end{tabularx}
\end{center}
\end{minipage}
\caption{List of generalized composite crystals without local valley polarization Crys\((2p,m/n,o)\) with filling fraction per layer per valley \(\nu_L\).}
\label{TableS4}
\end{center}
\end{table}

\subsection{Spontaneous Interlayer Coherent States}
We propose the generalization of spontaneous interlayer coherent (SILC) states \cite{PRB2002_Hanna, PRB2003_Schliemann, PRB2009_Burnell}, and compute their Coulomb energies per particle \(E_{\mathrm{Coulomb}}/N_{\mathrm{tot}}\) from \(H_{Coulomb}\) and their tunneling energies per particle \(E_{\mathrm{Tunnel}}/N_{\mathrm{tot}}\) from \(H_{\mathrm{Tunnel}}\) of the model Hamiltonian.

\paragraph{With Local Valley Polarization} Neglecting the valley indices, we define the SILC state to be
\begin{equation}
    \ket{\Psi_0} = \prod_q \prod_{m=-N_\phi/2}^{N_\phi/2} c_{q,m}^\dagger\ket{0},
\end{equation}
where \(N_\phi\) is the number of flux quanta, \(N_L\) is the number of layers in the system, \(\bar{\nu}\) is the filling fraction per layer, \(\alpha=d/2R\), \(R\) is the radius of the Haldane sphere, and
\begin{equation}
    c_{q,m} = \frac{1}{\sqrt{N_L}}\sum_{j=1}^{N_L} e^{iqjd} c_{j,m}, \quad \bra{\vartheta,\varphi}c_{j,m}\ket{0} = \sqrt{\frac{N_\phi+1}{2\pi {l_B}^2 N_\phi}\binom{N_\phi}{\frac{N_\phi}{2}+m}}\ u^{\frac{N_\phi}{2}+m}v^{\frac{N_\phi}{2}-m}.
\end{equation}
Then, the density matrix is given as
\begin{equation}
    \rho_{j_1,j_2}(m) = \bra{\Psi_0}c_{j_1,m}^\dagger c_{j_2,m}\ket{\Psi_0} 
    = \frac{1}{N_L}\sum_{l=1}^{N_L} e^{i\pi(j_1-j_2)\left(\frac{2l}{N_L}-1\right)} \mathbf{1}_{(-\bar{\nu},\bar{\nu}]}\left(\frac{2l}{N_L}-1\right) 
    \equiv \rho_0^{N_L}(|j_1-j_2|)
\end{equation}
with \(\mathbf{1}\) being the indicator function. Note that
\begin{equation}
    \rho_0^{\infty}(j) = \lim\limits_{N_L \to \infty} \rho^{N_L}(j) = \frac{\sin(\pi j \bar{\nu})}{\pi j}.
\end{equation}
Moreover, for \(r=2R\sin(\vartheta/2)\), the pair correlation function \cite{Book_Jain} is given as
\begin{align}
    g_j^{N_L}(r) &= \frac{1}{\rho^2}\bra{\Psi_0}\psi_0^\dagger(0,0)\psi_j^\dagger(\vartheta,0)\psi_j(\vartheta,0)\psi_0(0,0)\ket{\Psi_0} \nonumber \\
    &= \frac{1}{\bar{\nu}^2} \left[\left(\frac{1}{N_L}\sum_{l=1}^{N_L} \mathbf{1}_{(-\bar{\nu},\bar{\nu}]}\left(\frac{2l}{N_L}-1\right)\right)^2-\cos^{2 N_\phi}(\vartheta/2) \left(\rho_0^{N_L}(j)\right)^2\right],
\end{align}
where \(\psi_j(\vartheta,\varphi) = \sum_m \bra{\vartheta,\varphi}c_{j,m}\ket{0} c_{j,m}.\) That is, 
\begin{equation}
    g_j^{\infty}(r) = \lim\limits_{N_L \to \infty} g_j^{N_L}(r) = 1 - \cos^{2 N_\phi}(\vartheta/2) \left(\frac{\sin(\pi j \bar{\nu})}{\pi j \bar{\nu}}\right)^2.
\end{equation}
As a result, we obtain the Coulomb energy per particle of SILC state in the limit \(N_L\to\infty\):
\begin{align} \label{SILC1}
    \frac{E_{\mathrm{Coulomb}}}{N_{\mathrm{tot}}} &= -\frac{\rho e^2}{2\epsilon} \sum_{j=-\infty}^{\infty} \int \frac{1}{\sqrt{r^2+j^2 d^2}}\left(1 - g_j^\infty (r)\right) \mathrm{d}\mathbf{r} \nonumber \\
    &= -\frac{e^2}{\epsilon l_B}\cdot\frac{\bar{\nu}(N_\phi+1)}{2\sqrt{2N_\phi}} \sum_{j=-\infty}^{\infty} \left(\frac{\sin(\pi j \bar{\nu})}{\pi j \bar{\nu}}\right)^2 \int_0^1 \frac{2x}{\sqrt{x^2+j^2\alpha^2}}(1-x^2)^{N_\phi} \mathrm{d}x,
\end{align}
where \(e\) is the electric charge and \(\epsilon\) is the permittivity. Furthermore, with the valley indices intact, we rewrite the interlayer tunneling Hamiltonian:
\begin{equation}
    H_{\mathrm{Tunnel}} = -{t}_\perp\sum_{j,\sigma,m} c_{j,\sigma,m}^\dagger c_{j+1,\sigma,m} + (h.c.).
\end{equation}
That is, if the same valley is occupied for each layer, the tunneling energy per particle is 
\begin{align}
    \frac{E_{\mathrm{Tunnel}}}{N_{\mathrm{tot}}} = \lim\limits_{N_L\to\infty} \frac{1}{\bar{N} \cdot N_L}\bra{\Psi_0}H_{\mathrm{Tunnel}}\ket{\Psi_0}
    = -\frac{2\tilde{t}_\perp \rho_0^\infty(1)}{\bar{\nu}} = -2\tilde{t}_\perp \frac{\sin(\pi \bar{\nu})}{\pi \bar{\nu}}.
\end{align}
with \(\bar{N}\) being the number of particles per layer. Meanwhile, \(E_{\mathrm{Tunnel}}=0\) if the valley texture is layer-alternating.

\paragraph{Without Local Valley Polarization} Similarly, we define the SILC state with full valley indices as 
\begin{equation}
    \ket{\Psi_0} = \prod_q \prod_{\sigma} \prod_{m=-N_\phi/2}^{N_\phi/2} c_{q,\sigma, m}^\dagger\ket{0},
\end{equation}
where \(\nu_L\) is the given filling fraction per layer per valley, and
\begin{equation}
    \quad c_{q,\sigma, m} = \frac{1}{\sqrt{N_L}}\sum_{j=1}^{N_L} e^{iqjd} c_{j,\sigma,m}, \quad \bra{\vartheta,\varphi}c_{j,\sigma,m}\ket{0} = \sqrt{\frac{N_\phi+1}{2\pi {l_B}^2 N_\phi}\binom{N_\phi}{\frac{N_\phi}{2}+m}}\ u^{\frac{N_\phi}{2}+m}v^{\frac{N_\phi}{2}-m}.
\end{equation}
Then, the density matrix is given as 
\begin{align}
    \rho_{j_1,j_2}(m) & = \sum_{\sigma_1,\sigma_2} \bra{\Psi_0} c_{j_1,\sigma_1,m}^\dagger c_{j_2,\sigma_2 m} \ket{\Psi_0} \nonumber \\ & = \sum_{\sigma_1,\sigma_2} \delta_{\sigma_1\sigma_2} \cdot \frac{1}{N_L}\sum_{l=1}^{N_L} e^{i\pi(j_1-j_2)\left(\frac{2l}{N_L}-1\right)} \mathbf{1}_{(-\nu_L,\nu_L]}\left(\frac{2l}{N_L}-1\right) \equiv {\rho}_0^{N_L}(|j_1-j_2|),
\end{align}
therefore
\begin{equation}
    \rho_0^{\infty}(j) = \lim\limits_{N_L \to \infty} \rho^{N_L}(j) = \frac{2\sin(\pi j \nu_L)}{\pi j}.
\end{equation}
Moreover, letting \(\psi_{j,\sigma}(\vartheta,\varphi) = \sum_m \bra{\vartheta,\varphi}c_{j,\sigma,m}\ket{0} c_{j,\sigma,m}\), the pair correlation function is given as
\begin{align}
    g_j^{N_L}(r)
    &= \sum_{\sigma,\sigma'} \frac{1}{\rho^2}\bra{\Psi_0}\psi_{0,\sigma}^\dagger(0,0)\psi_{j,\sigma'}^\dagger(\vartheta,0)\psi_{j,\sigma'}(\vartheta,0)\psi_{0,\sigma}(0,0)\ket{\Psi_0} \nonumber \\
    &= \frac{1}{4\nu_L^2} \left[\left(\frac{2}{N_L}\sum_{l=1}^{N_L} \mathbf{1}_{(-\nu_L,\nu_L]}\left(\frac{2l}{N_L}-1\right)\right)^2- 2\cos^{2 N_\phi}(\vartheta/2) \left(\frac{\rho_0^{N_L}(j)}{2}\right)^2\right],
\end{align}
thus
\begin{equation}
    g_j^{\infty}(r) = \lim\limits_{N_L \to \infty} g_j^{N_L}(r) = 1 - \frac{1}{2}\cos^{2 N_\phi}(\vartheta/2) \left(\frac{\sin(\pi j \nu_L)}{\pi j \nu_L}\right)^2.
\end{equation}
As a result, in the limit \(N_L\to\infty\), the Coulomb energy per particle is
\begin{align} \label{SILC2}
    \frac{E_{\mathrm{Coulomb}}}{N_{\mathrm{tot}}} &= -\frac{\rho e^2}{2\epsilon} \sum_{j=-\infty}^{\infty} \int \frac{1}{\sqrt{r^2+j^2 d^2}}\left(1 - g_j^\infty (r)\right) \mathrm{d}\mathbf{r} \nonumber \\
    &= -\frac{e^2}{\epsilon l_B}\cdot\frac{\nu_L(N_\phi+1)}{2\sqrt{2N_\phi}} \sum_{j=-\infty}^{\infty} \left(\frac{\sin(\pi j \nu_L)}{\pi j \nu_L}\right)^2 \int_0^1 \frac{2x}{\sqrt{x^2+j^2\alpha^2}}(1-x^2)^{N_\phi} \mathrm{d}x.
\end{align}
We rewrite the tunneling Hamiltonian as
\begin{equation}
    H_{\mathrm{Tunnel}} = -t_\perp \sum_{j,m,\sigma} c_{j,\sigma,m}^\dagger c_{j+1,\sigma,m} + (h.c.).
\end{equation}
Since \(\bra{\Psi_0} c_{j,\sigma,m}^\dagger c_{j+1,-\sigma,m} \ket{\Psi_0} = 0\) for every valley \(\sigma\),
\begin{align}
    \frac{E_{\mathrm{Tunnel}}}{N_{\mathrm{tot}}} = \lim\limits_{N_L\to\infty} \frac{1}{\bar{N} \cdot N_L}\bra{\Psi_0}H_{\mathrm{Tunnel}}\ket{\Psi_0}
    = -\frac{2t_\perp\rho_0^\infty(1)}{\bar{\nu}}=-2t_\perp \frac{\sin(\pi \nu_L)}{\pi \nu_L}.
\end{align}

\paragraph{Numerical Calculation of Coulomb Energies}
Letting \(\delta = d/l_B\), so that \(j^2\alpha^2 = \frac{j^2\delta^2}{2N_\phi}\), we calculate the integral in \(E_{\mathrm{Coulomb}}/N_{\mathrm{tot}}\) to rewrite Eqs.~\eqref{SILC1},~\eqref{SILC2}:
\begin{equation}
    \int_0^1 \frac{2x}{\sqrt{x^2+j^2\alpha^2}}(1-x^2)^{N_\phi} \mathrm{d}x =
    \begin{cases}
        \sqrt{\pi} \frac{\Gamma(N_\phi+1)}{\Gamma(N_\phi+3/2)} & \quad \text{if } j\delta = 0\\
        \sqrt{2N_\phi} N_\phi \frac{\Gamma(N_\phi)}{\Gamma(N_\phi+2)} \frac{_2F_1(1/2,1,N_\phi+2;\frac{-2N_\phi}{j^2\delta^2})}{|j\delta|} & \quad \text{otherwise,}
    \end{cases}
\end{equation}
where \(\Gamma\) is the gamma function, and \(_2F_1\) is the Gauss's hypergeometric function. Infinite sums are evaluated by the mpmath package \cite{mpmath}. Richardson extrapolation, Shanks extrapolation, and Euler-Maclaurin summation formula are used. To estimate the Coulomb energies of SILC states, we assume that the number of particles per each layer is set to be \(\bar{N} = 48\). That is, for the valley-coherent state, the number of particles per each layer per each valley is \(N = 24\). For each filling per layer \(\bar{\nu}\), \(N_\phi = \bar{N}/\bar{\nu} - 1\), or equivalently, \(N_\phi = N/\nu_L - 1\).

\subsection{Product of States}
Let \(A\oplus B\) be the product of states \(A\) and \(B\), so that its wavefunction is given as \(\Psi_{A\oplus B} = \Psi_A \cdot \Psi_B\) where \(\Psi_{A(B)}\) are the wavefunctions of \(A(B)\). Its physical meaning that \(A\) and \(B\) coexist without any coherence between them. In the language of K-matrix formalism \cite{Book_Wen}, if \(A(B)\) corresponds to \((K_{A(B)}, \mathbf{t}_{A(B)}, \mathbf{s}_{A(B)})\), \(A\oplus B\) corresponds to \((K_A \oplus K_B, \mathbf{t}_A \oplus \mathbf{t}_B, \mathbf{s}_A \oplus \mathbf{s}_B)\), where
\begin{equation} \label{kmatofcoprod}
	K_A \oplus K_B  = \begin{pmatrix} K_A & \mathbf{0} \\ \mathbf{0} & K_B \end{pmatrix},\ \mathbf{t}_A \oplus \mathbf{t}_B = \begin{pmatrix} \mathbf{t}_A \\ \mathbf{t}_B \end{pmatrix},\ \mathbf{s}_A \oplus \mathbf{s}_B = \begin{pmatrix} \mathbf{s}_A \\ \mathbf{s}_B \end{pmatrix}. 
\end{equation}
Then, if \(\nu_{A(B)}\) and \(S_{A(B)}\) is the filling fraction and the shift of \(A(B)\), the filling fraction and the shift of \(A\oplus B\) is given as
\begin{equation}\label{fillingshift}
	\nu_{A\oplus B} = \nu_A + \nu_B, \quad S_{A\oplus B} = \frac{\nu_A S_A + \nu_B S_B}{\nu_A + \nu_B}.
\end{equation}
We derive below relations to compute the energies of product states.

\paragraph{Relation for the Products of Same States}
For any uniform density state \(X\), 	
\begin{equation}
	\frac{E_{\mathrm{Coulomb}}^{X\oplus X}}{N_{X\oplus X}} = \frac{E_{\mathrm{Coulomb}}^X}{N_X}.
\end{equation}
From the above, it is straightforward that Coulomb energy per particle for any generalized Halperin liquid \((m000)\) is equal to the Coulomb energy per particle of the single-component \(1/m\) Laughlin state. Similarly, the Coulomb energy per particle for any generalized Halperin liquid \((mn/00)\) is equal to the Coulomb energy per particle of the usual valley-coherent \((mnm)\) Halperin state.

\paragraph{Relation for the Products of Different States}
Let \(E_{\mathrm{Coulomb}}^{X}(N)\) be the Coulomb energy of the state \(X\) consisting \(N\) particles, whose filling fraction is \(\nu_X\) and the shift is \(S_X\). For any uniform states \(X_1\) and \(X_2\), 
\begin{align}
	\frac{E_{\mathrm{Coulomb}}^{X_1\oplus X_2}(N)}{N} & \approx \frac{N-\nu_{X_1}S_{X_1}-\nu_{X_2}S_{X_2}}{N-\nu_{X_1}S_{X_1}}\frac{\nu_{X_1}}{\nu_{X_1}+\nu_{X_2}}\frac{E_{\mathrm{Coulomb}}^{X_1}(N)}{N} \nonumber \\
	& + \frac{N-\nu_{X_1}S_{X_1}-\nu_{X_2}S_{X_2}}{N-\nu_{X_2}S_{X_2}}\frac{\nu_{X_2}}{\nu_{X_1}+\nu_{X_2}}\frac{E_{\mathrm{Coulomb}}^{X_2}(N)}{N}.
\end{align}

\subsection{Particle-Hole Conjugate States}
Let \(X\) be the state with filling \(\nu_X\) and shift \(S_X\), whose K-matrix description is \((K_X,\mathbf{t}_X,\mathbf{s}_X)\). Define \(\bar{X}\) to be the particle-hole conjugate of \(X\). Then, \(\bar{X}\) is characterized by \cite{Book_Wen}
\begin{equation}\label{kmatphconj}
	K_{\bar{X}} = \begin{pmatrix} \mathbf{1} & \mathbf{0} \\ \mathbf{0} & -K_X \end{pmatrix},\ \mathbf{t}_{\bar{X}}  = \begin{pmatrix} \mathbf{1} \\ -\mathbf{t}_X \end{pmatrix},\ \mathbf{s}_{\bar{X}} = \begin{pmatrix} \mathbf{\frac{1}{2}} \\ -\mathbf{s}_X \end{pmatrix}.
\end{equation}
Note that the above characterization corresponds to a valid topological order. Moreover, the filling fraction and the shift of the state \(\bar{X}\) is given by
\begin{equation}
	\nu_{\bar{X}} = 1 - \nu_X, \quad S_{\bar{X}} = \frac{1 - \nu_X S_X}{1 - \nu_X}.
\end{equation}
We derive the below relation to compute the energies of particle-hole conjugate states.

\paragraph{Relation for Particle-Hole Conjugate States} Let \(E_{\mathrm{Coulomb}}^{X}(N)\) be the Coulomb energy of the state \(X\) consisting \(N\) particles, whose filling fraction is \(\nu_X\) and the shift is \(S_X\). Similarly, let \(E_{\mathrm{Coulomb}}^{\bar{X}}(N)\) and \(E_{\mathrm{Coulomb}}^{\mathbf{1}}(N)\) be the energies of the particle-hole conjugate of \(X\) and the one-filled (lowest) Landau level consisting \(N\) particles each. Then,
\begin{equation}	
	\frac{E_{\mathrm{Coulomb}}^{\bar{X}}(N)}{N} \approx \frac{\nu_X}{1-\nu_X}\frac{N-1+\nu_X S_X}{N-\nu_X S_X}\frac{E_{\mathrm{Coulomb}}^{X}(N)}{N} + \frac{1-2\nu_X}{1-\nu_X}\frac{N-1+\nu_X S_X}{N-1}\frac{E_{\mathrm{Coulomb}}^{\mathbf{1}}(N)}{N}.
\end{equation}	

\subsection{Staged States}
We refer a state whose charge distribution varies along the stacking direction as staged state. Such non-uniformity leads to an energy compensation of the system, which is referred as the staging energy \(E_{\mathrm{Staging}}\) \cite{PRB1989_Qiu, PRB2009_Burnell}. In particular, we consider four different staging patterns and derive their staging energies.

\paragraph{Staging Pattern 1} Given \(n\in\mathbb{N}\), the system satisfies
\begin{equation}
    \nu_l - \overline{\nu} =
    \begin{cases}
    n \delta\nu     & \quad \text{if } l = k(n+1) \text{ for some } k\in\mathbb{Z}\\
    - \delta\nu     & \quad \text{otherwise}
  \end{cases} 
\end{equation}
for all \(l\in\mathbb{Z}\), where \(\nu_l\) is the filling fraction in layer \(l\), and \(\overline{\nu}\) is the average filling fraction per layer. Assuming \(d \ll R\), where \(R\) is the radius of the Haldane sphere, the staging energy per particle is given by
\begin{equation}
    \frac{E_{\mathrm{Staging}}}{N_{\mathrm{tot}}} \approx \frac{e^2}{\epsilon l_B} \frac{d}{l_B} \frac{\delta\nu^2 n(n+2)}{12\bar{\nu}}.
\end{equation}

\paragraph{Staging Pattern 2} Given \(n\in\mathbb{N}\), the system satisfies
\begin{equation}
    \nu_l - \overline{\nu} =
    \begin{cases}
    \left(n+\frac{1}{2}\right) \delta \nu  & \; \text{if }\exists k\in\mathbb{Z} \text{ s.t. } l = k(2n+3) \text{ or } l = n+1 + k(2n+3) \\
    - \delta \nu  & \; \text{otherwise},
  \end{cases}
\end{equation}
for each \(l\in\mathbb{Z}\). Assuming \(d \ll R\), the staging energy per particle is given by
\begin{equation}
    \frac{E_{\mathrm{Staging}}}{N_{\mathrm{tot}}} \approx \frac{e^2}{\epsilon l_B} \frac{d}{l_B} \frac{\delta\nu^2 (n+1)(n+2)}{12\bar{\nu}}.
\end{equation}

\paragraph{Staging Pattern 3} Given \(n\in\mathbb{N}\), the system satisfies
\begin{equation}
    \nu_l - \overline{\nu} =
    \begin{cases}
    \frac{n}{2} \delta \nu  & \quad \text{if } \exists k\in\mathbb{Z} \text{ s.t. } l = k(n+2) \text{ or } l = 1 + k(n+2) \\
    - \delta \nu  & \quad \text{otherwise},
  \end{cases}
\end{equation}
for each \(l\in\mathbb{Z}\). Assuming \(d \ll R\), the staging energy per particle is given by
\begin{equation}
    \frac{E_{\mathrm{Staging}}}{N_{\mathrm{tot}}} \approx \frac{e^2}{\epsilon l_B} \frac{d}{l_B} \frac{\delta\nu^2 n(n+1)}{12\bar{\nu}}.
\end{equation}

\paragraph{Staging Pattern 4} Given \(n\in\mathbb{N}\), the system satisfies
\begin{equation} 
    \nu_l - \overline{\nu} =
    \begin{cases}
    \left(\frac{3n\pm 1}{3}\right) \delta \nu  &  \text{if }\exists k\in\mathbb{Z} \text{ s.t. } l = k(3n+3\pm 1), l = n+1+k(3n+3\pm 1) \text{ or } l = 2n+2\pm 1+k(3n+3\pm 1) \\
    - \delta \nu  &  \text{otherwise},
  \end{cases}
\end{equation}
for each \(l\in\mathbb{Z}\). Assuming \(d \ll R\), the staging energy per particle is given by
\begin{equation}
    \frac{E_{\mathrm{Staging}}}{N_{\mathrm{tot}}} \approx \frac{e^2}{\epsilon l_B} \frac{d}{l_B} \frac{\delta\nu^2 (n+1)(3n+3\pm 2)}{36\bar{\nu}}.
\end{equation}

\supnotesection{Phase Diagrams for Three-Dimensional Graphene Multilayers}

\paragraph{Model Parameters} \ref{TableS5} provides the model parameters for the phase diagram in the main text (Fig.~3). Interlayer tunneling strengths and the phenomenological parameters \(V_1\), \(V_2\), and \(V_3\) are estimated from the literature.

\begin{table}[H]
\begin{center}
\begin{tabular}{c || c | c | c}
	\hline 
	& Bernal graphite  & ATG multilayer \cite{PRL2024_Lu} & HTG multilayer \\ \hline\hline
	\(d\) (\AA) & 3.35 \cite{PRL2024_Lu} & 3.35 \cite{PRL2024_Lu} & 3.35 \cite{PRL2024_Lu} \\ \hline
   \(\epsilon_r\) & 12.5 \cite{PRB2009_Burnell} & 12.5 \cite{PRB2009_Burnell} & 12.5 \cite{PRB2009_Burnell} \\
   \hline
   \(W\) (meV) & 40 \cite{PRL2007_Bernevig,PRB2009_Burnell} & 7.12 \cite{PRL2007_Santos,PRL2024_Lu} & 4.36\\
   \hline
    \(t_\perp\) (meV) & 10 & 1.78 & 1.09\\
   \hline
    \(t_\perp\times l_B/d\) (\(e^2/{\epsilon l_B}\)) & 0.0023 & 0.00041 & 0.00025  \\
   \hline\hline
   \(V_1 \times l_B/d\) (\(e^2/\epsilon l_B\)) & 0.0101493 & 0.0101493 \cite{NatComm2025_Kim} & 0.0101493  \\ \hline
   \(V_2 \times l_B/d\) (\(e^2/\epsilon l_B\)) & 0.507463 & 0.40597 \cite{NatComm2025_Kim} & 0.314627  \\ \hline
   \(V_3 \times l_B/d\) (\(e^2/\epsilon l_B\)) & 0.101493 & 0.202985 \cite{NatComm2025_Kim} & 0.294328  \\ \hline
\end{tabular}
\caption{Interlayer distance \(d\), relative dielectric constant \(\epsilon_r\), bandwidth \(W\), interlayer tunneling strength \(t_\perp\), and phenomenological parameters \(V_1\), \(V_2\), \(V_3\) of three-dimensional Bernal graphite, alternating twisted graphene (ATG) multilayer, and helical twisted graphene (HTG) multilayer in the main text.}
\label{TableS5}
\end{center}
\end{table}

\begin{figure}[H]
\begin{center}
\includegraphics[width=1\textwidth]{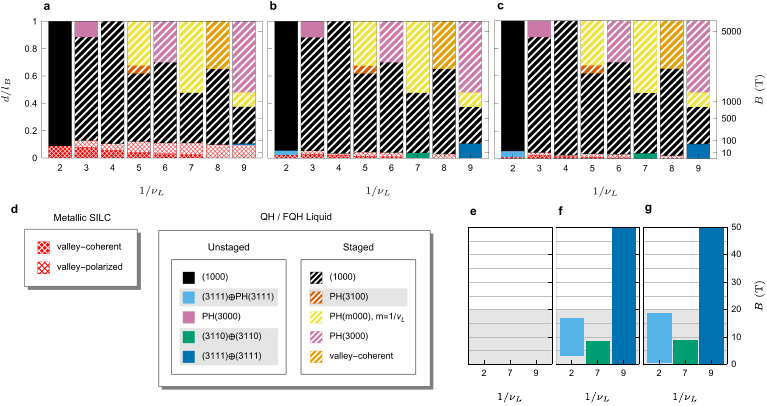}
\caption{{\bf Phase diagrams with different phenomenological parameters.} Phenomenological parameters are provided in  \ref{TableS6}. {\bf a,} Phase diagram for three-dimensional Bernal graphite (\(t_\perp \approx 10 \text{ meV}\)), {\bf b,} alternating twisted graphene multilayer (\(t_\perp \approx 1.78 \text{ meV}\)), and {\bf c,} helical twisted graphene multilayer (\(t_\perp \approx 1.09 \text{ meV}\)). {\bf d,} Phase labels. Interlayer-coherent FQH phases are highlighted in gray. Valley ordering and entanglement patterns are provided in~\ref{figS1} and~\ref{figS2}. {\bf e--g,} Parameter regimes in which generalized Halperin states emerge at low magnetic fields for {\bf e,} Bernal graphite, {\bf f,} alternating twisted graphite, and {\bf g,} helical twisted graphite. Experimentally accessible range of magnetic fields is highlighted in gray.}
\label{figS3}
\end{center}
\end{figure}

\paragraph{Stability of the Phase Diagram} \ref{TableS6} provides phenomenological parameters \(V_1\), \(V_2\), and \(V_3\) for \ref{figS3}, much smaller than those for Fig.~3 in the main text. We adopt the same interlayer tunneling strengths from \ref{TableS5}.

\begin{table}[H]
\begin{center}
\begin{tabular}{c || c | c | c}
	\hline 
	& Bernal graphite  & ATG multilayer & HTG multilayer \\ 
   \hline\hline
   \(V_1 \times l_B/d\) (\(e^2/\epsilon l_B\)) & 0.00101493 & 0.00101493 & 0.00101493  \\ \hline
   \(V_1 \times l_B/d\) (\(e^2/\epsilon l_B\)) & 0.00304478 & 0.00253731 & 0.00162388  \\ \hline
   \(V_3 \times l_B/d\) (\(e^2/\epsilon l_B\)) & 0.00202985 & 0.00152239 & 0.0014209  \\ \hline
\end{tabular}
\caption{Phenomenological parameters \(V_1\), \(V_2\), \(V_3\) of three-dimensional Bernal graphite, alternating twisted graphene (ATG) multilyer, and helical twisted graphene (HTG) multilayer for \ref{figS3}.}
\label{TableS6}
\end{center}
\end{table}

The phase diagrams in \ref{figS3} exhibit the same significant features with those in Fig.~3. First, in the strong-field regime, the phase diagrams have no dependence on the twist angle, and are dominated by the staged liquids (\ref{figS3}a--c). Meanwhile, under the magnetic field of experimentally accessible strength, the twist angle has a huge impact. Without any twist, metallic states dominate the phase diagram (\ref{figS3}e). However, alternating and helical twisted graphene multilayers show robust generalized Halperin states at \(\nu_L=1/2\), \(\nu_L=1/7\) and \(\nu_L=1/9\). As a matter of fact, these states are a bit more favored in \ref{figS3} than in Fig.~3. As a result, we emphasize that the phase diagrams remain stable against the variations of the phenomenological model parameters. In particular, the emergence of the generalized Halperin states are far from being sensitive, supporting the possible observations of these exotic phases in experiments.

\begin{figure}[H]
\begin{center}
\includegraphics[width=0.9\textwidth]{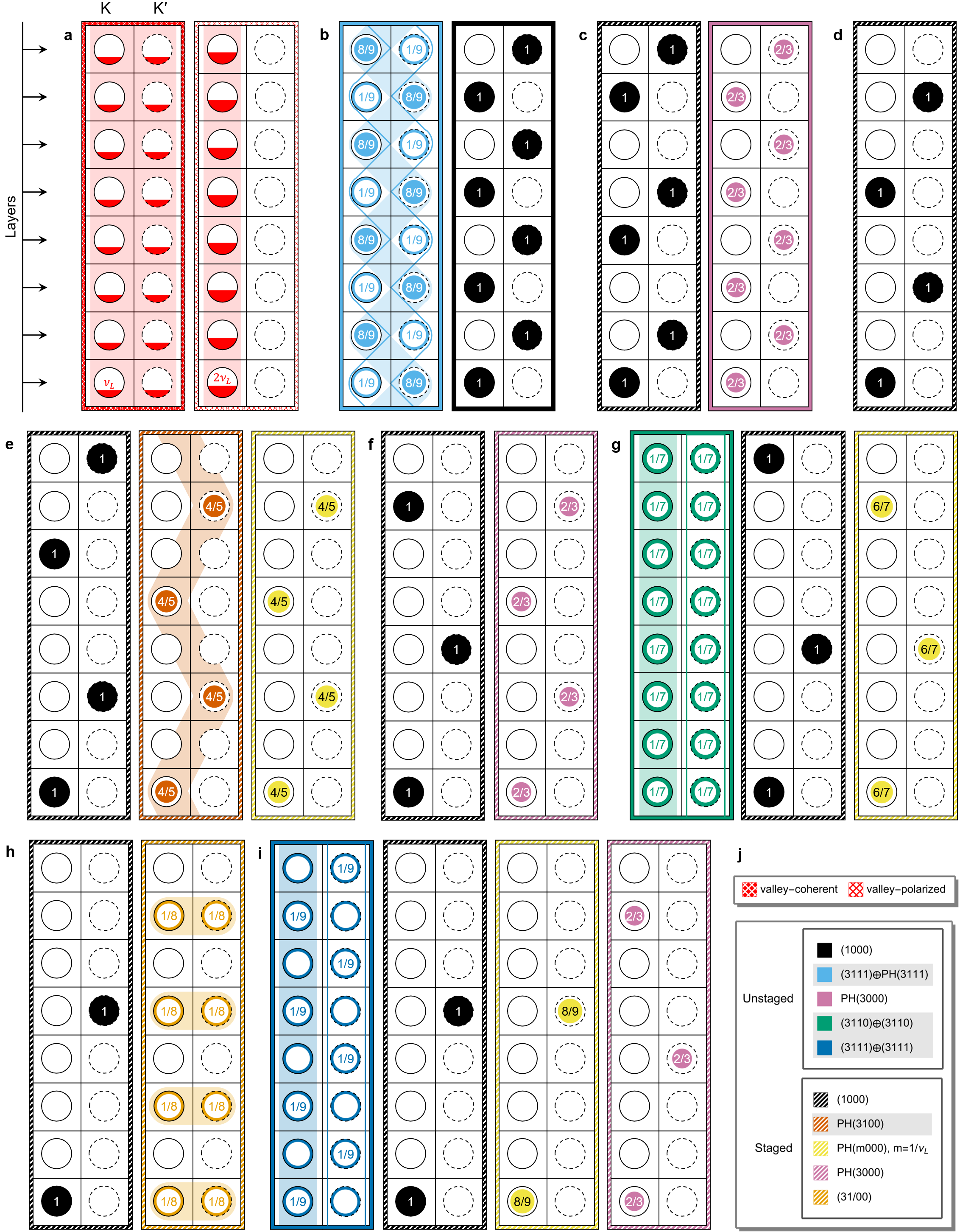}
\caption{{\bf Schematics of valley ordering and entanglement patterns of Bernal graphite and alternating twisted graphene multilayer.} {\bf a,} Symmetry-breaking and entanglement patterns of metallic spontaneous interlayer coherent states with filling fraction per layer per valley \(\nu_L\), and {\bf b,} quantum Hall states with \(\nu_L=1/2\), {\bf c,} \(\nu_L=1/3\), {\bf d,} \(\nu_L=1/4\), {\bf e,} \(\nu_L=1/5\), {\bf f,} \(\nu_L=1/6\), {\bf g,} \(\nu_L=1/7\), {\bf h,} \(\nu_L=1/8\), {\bf i,} \(\nu_L=1/9\) in Fig.~2a,b and \ref{figS3}b,c. Each state is indicated by its frame. {\bf j,} Phase labels. Interlayer-coherent fractional quantum Hall phases are highlighted in gray.}
\label{figS1}
\end{center}
\end{figure}

\begin{figure}[H]
\begin{center}
\includegraphics[width=0.9\textwidth]{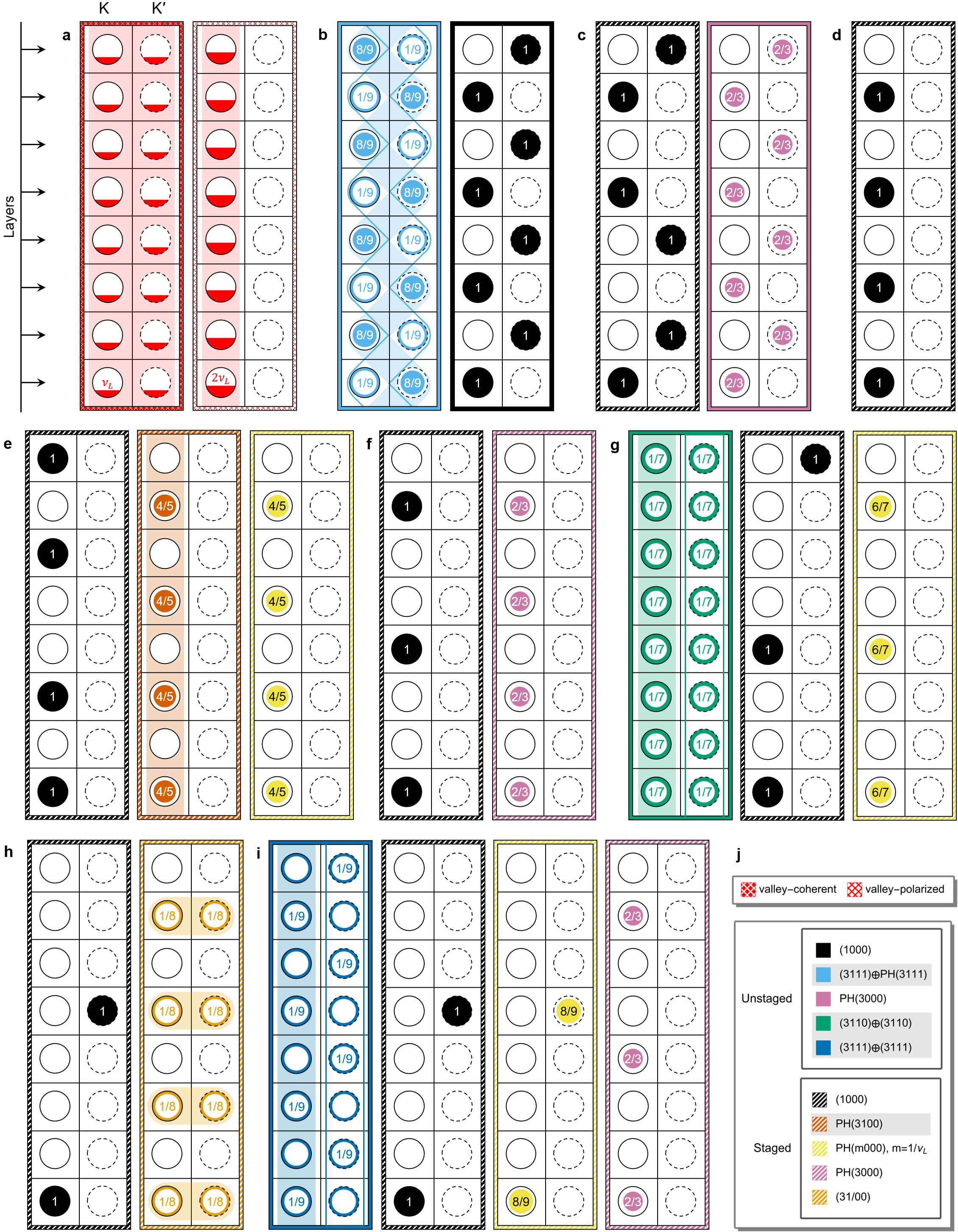}
\caption{{\bf Valley ordering and entanglement patterns of helical twisted graphene multilayer.} The twist angle is $10^{\circ} \ll \theta < 30^{\circ}$. {\bf a,} Symmetry-breaking and entanglement patterns of metallic spontaneous interlayer coherent states with filling fraction per layer per valley \(\nu_L\), and {\bf b,} quantum Hall states with \(\nu_L=1/2\), {\bf c,} \(\nu_L=1/3\), {\bf d,} \(\nu_L=1/4\), {\bf e,} \(\nu_L=1/5\), {\bf f,} \(\nu_L=1/6\), {\bf g,} \(\nu_L=1/7\), {\bf h,} \(\nu_L=1/8\), {\bf i,} \(\nu_L=1/9\) in Figs.~2c and \ref{figS3}c. Each state is indicated by its frame. {\bf j,} Phase labels. Interlayer-coherent fractional quantum Hall phases are highlighted in gray.}
\label{figS2}
\end{center}
\end{figure}

\paragraph{Valley Ordering and Entanglement Patterns} We provide full schematics of valley ordering and entanglement patterns of all phases in Fig.~3 and \ref{figS3}. \ref{figS1} shows the symmetry-breaking and entanglement patterns of the Bernal graphite and alternating twisted graphene multilayer. Symmetry-breaking and entanglement patterns of helical twisted graphene multilayer are illustrated in \ref{figS2}. We note that, in contrast with the Bernal graphite and alternating twisted graphene multilayer, the valley textures of helical twisted multilayer are dependent on the twist angle \(\theta\). Here, we consider \(\theta\) slightly less than \(30^\circ\).

\supnotesection{Phase Diagrams for Three-Dimensional Multilayer Transition Metal Dichalcogenides}

For the transition metal dichalcogenides (TMDs), whose valley degeneracy is also lifted due to spin-valley locking \cite{PRRes2020_Xuan}, we use \(H = H_{Coulomb} + H_{Tunnel}\) while neglecting the valley indices. 

\paragraph{Model Parameters} \ref{TableS7} provides the interlayer tunneling strengths of the 2H-\(\mathrm{MX_2}\) TMDs estimated from the literature. Meanwhile, \ref{TableS8} provides the possible values interlayer tunneling strengths of three-dimensional twisted multilayer TMDs.

\begin{table}[H]
\begin{center}
\begin{tabular}{c || c | c | c | c | c  }
	\hline 
	& \(\mathrm{WS_2}\) & \(\mathrm{WSe_2}\) & \(\mathrm{MoS_2}\) & \(\mathrm{MoSe_2}\) & \(\mathrm{MoTe_2}\) \\
	\hline \hline
	\(d\) (\AA) & 6.229 \cite{ACS2022_Maduro} & 6.48 \cite{PRB1997_Finteis} & 6.1475 
	\cite{PRB35_Coehoorn} & 6.474 \cite{PSS2017_Gusakova} & 6.789 \cite{PRMat2018_Pike} \\
   \hline
   \(\epsilon_r\) & 14.4 \cite{NPJ2018_Laturia} & 15.9 \cite{NPJ2018_Laturia} & 15.9 \cite{NPJ2018_Laturia} & 17.7 \cite{NPJ2018_Laturia} & 21.9 \cite{NPJ2018_Laturia} \\
   \hline
   \(W\) (meV) & 25.27 \cite{ACS2022_Maduro} & 40 \cite{PRB1997_Finteis} & 50 \cite{SR2014_Chang} & 68.59 \cite{NatComm2026_Alarab} & 59.6 \cite{MRL2020_Oliva} \\
   \hline
    \(t_\perp\) (meV) & 6.32 & 10 \cite{PRB1997_Finteis} & 12.5 \cite{SR2014_Chang} & 17.15 \cite{NatComm2026_Alarab} & 14.9 \cite{MRL2020_Oliva} \\
   \hline
   \(t_\perp\times l_B/d\) (\(e^2/{\epsilon l_B}\)) & 0.00313 & 0.0057 & 0.0068 & 0.0109 & 0.0122 \\
   \hline 
\end{tabular}
\caption{Interlayer distance \(d\), relative dielectric constant \(\epsilon_r\), bandwidth \(W\), and interlayer tunneling strength \(t_\perp\) of bulk 2H-\(\mathrm{MX_2}\) transition metal dichalcognides.}
\label{TableS7}
\end{center}
\end{table}

\begin{table}[H]
\begin{center}
\begin{tabular}{c || c | c | c | c | c  }
	\hline 
	\(t_\perp\times l_B/d\) (\(e^2/{\epsilon l_B}\)) & \(\mathrm{WS_2}\) & \(\mathrm{WSe_2}\) & \(\mathrm{MoS_2}\) & \(\mathrm{MoSe_2}\) & \(\mathrm{MoTe_2}\) \\
	\hline \hline
	0.0015  & 3.026 & 2.634 & 2.777 & 2.369 & 1.826 \\
    \hline
    0.00075 &  1.513 & 1.317 & 1.388 & 1.184 & 0.9128 \\
    \hline
    0.00015 & 0.3026 & 0.2634 & 0.2777 & 0.2369 & 0.1826\\
    \hline\hline
\end{tabular}
\caption{Estimated interlayer tunneling strength \(t_\perp\) in \(\mathrm{meV}\) for twisted transition metal dichalcogenides (TMDs) (\ref{figS4}f--h,j--l). Fixing \(t_\perp\times l_B/d\), \(t_\perp\) varies with the atomic species of TMDs. \(t_\perp\times l_B/d\) are considered to cover the possible range of interlayer tunneling strength in various types of alternating and helical twisted TMDs. \(t_\perp\times l_B/d = 0.0015\) (\ref{figS4}f,j),
\(t_\perp\times l_B/d = 0.00075\) (\ref{figS4}g,k), and \(t_\perp\times l_B/d = 0.00015\) (\ref{figS4}h,l) correspond to \(2~\mathrm{meV}\lesssim t_\perp \lesssim 3~\mathrm{meV}\), \(1~\mathrm{meV}\lesssim t_\perp \lesssim 1.5~\mathrm{meV}\), and \(0.2~\mathrm{meV}\lesssim t_\perp \lesssim 0.3~\mathrm{meV}\), respectively.}
\label{TableS8}
\end{center}
\end{table}

\begin{figure}[H]
\begin{center}
\includegraphics[width=1\textwidth]{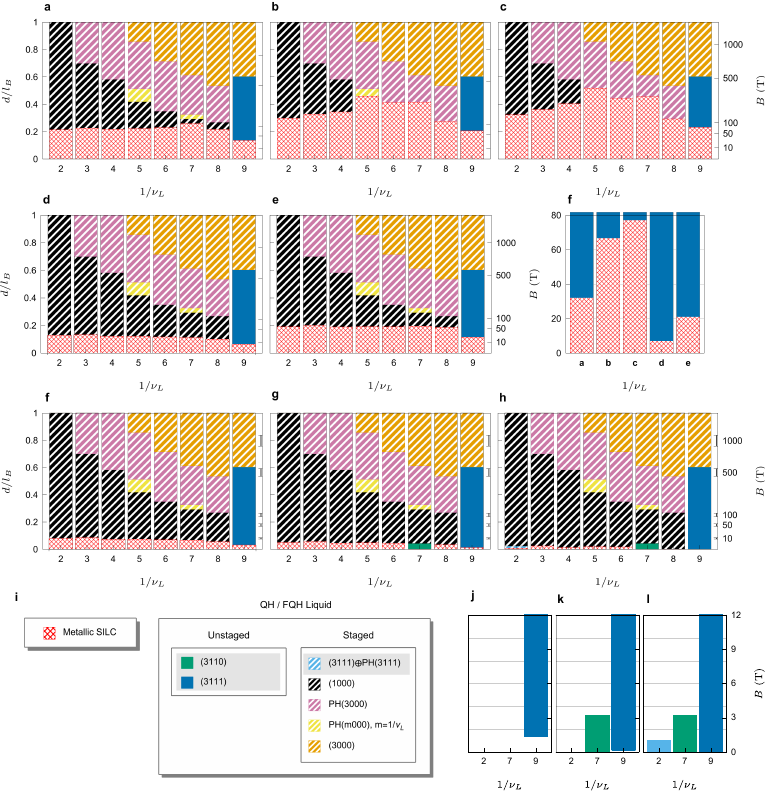}
\caption{{\bf Phase diagrams for transition metal dichalcogenides.} {\bf a,} Phase diagram for three-dimensional 2H-\(\mathrm{MX_2}\) transition metal dichalcogenides (TMDs) \(\mathrm{MoS_2}\) (\(t_\perp \approx 12.5 \text{ meV}\)), {\bf b,} \(\mathrm{MoSe_2}\) (\(t_\perp \approx 17.75 \text{ meV}\)), {\bf c,} \(\mathrm{MoTe_2}\) (\(t_\perp \approx 14.9 \text{ meV}\)), {\bf d,} \(\mathrm{WS_2}\) (\(t_\perp \approx 6.32 \text{ meV}\)), and {\bf e,} \(\mathrm{WSe_2}\) (\(t_\perp \approx 10 \text{ meV}\)). {\bf f,} Comparison of phase transitions from metallic state to generalized Halperin state at \(\nu_L=1/9\) among \(\mathrm{MoS_2}\) ({\bf a}), \(\mathrm{MoSe_2}\) ({\bf b}), \(\mathrm{MoTe_2}\) ({\bf c}), \(\mathrm{WS_2}\) ({\bf d}), and \(\mathrm{WSe_2}\) ({\bf e}), against the magnetic field. {\bf f,} Phase diagram of twisted TMD multilayers with \(2~\mathrm{meV}\lesssim t_\perp\lesssim 3~\mathrm{meV}\), {\bf g,} \(1~\mathrm{meV}\lesssim t_\perp\lesssim 1.5~\mathrm{meV}\), and {\bf h,} \(0.2~\mathrm{meV}\lesssim t_\perp\lesssim 0.3~\mathrm{meV}\). {\bf i,} Phase labels. Interlayer-coherent FQH phases are highlighted in gray. {\bf j,} Parameter regimes in which generalized Halperin states emerge at low magnetic fields for twisted TMD multilayers with \(2~\mathrm{meV}\lesssim t_\perp\lesssim 3~\mathrm{meV}\), {\bf k,} \(1~\mathrm{meV}\lesssim t_\perp\lesssim 1.5~\mathrm{meV}\), and {\bf l,} \(0.2~\mathrm{meV}\lesssim t_\perp\lesssim 0.3~\mathrm{meV}\).}
\label{figS4}
\end{center}
\end{figure}

\paragraph{Phase Diagrams} \ref{figS4} shows the phase diagrams for various TMDs of different atomic species and structures. We find that TMDs are the graphites, but without the valley degeneracy. First, in the strong-field regime, we observe decoupled staged quantum Hall states of two-dimensional Laughlin states or their particle-hole conjugates at every filling fraction per layer \(\nu_L\) of our interest (\ref{figS4}a--e,f--h). Such observation does not differ by the interlayer tunneling, or equivalently, the twist angle. Second, under low magnetic fields, metallic SILC states dominate the phase diagrams of three-dimensional TMDs without twist. (\ref{figS4}a--e). However, the critical magentic fields of the transitions from the SILC states to the liquid states differ by the atomic species. Among five different non-twisted TMDs, only the phase transition to the generalized Halperin state at \(\nu_L=1/9\) in \(\mathrm{WS_2}\) occurs at experimentally accessible range of magnetic field (\ref{figS4}f). However, this result is not much practical due to the electronic structure of the bulk 2H-\(\mathrm{WS_2}\). Meanwhile, note that the twisted multilayer TMDs host robust generalized Halperin states at \(\nu_L=1/2\), \(\nu_L=1/7\) and \(\nu_L=1/9\) in the experimentally accessible range of magnetic field, as twisted graphene multilayers do. 

\begin{figure}[H]
\begin{center}
\includegraphics[width=1\textwidth]{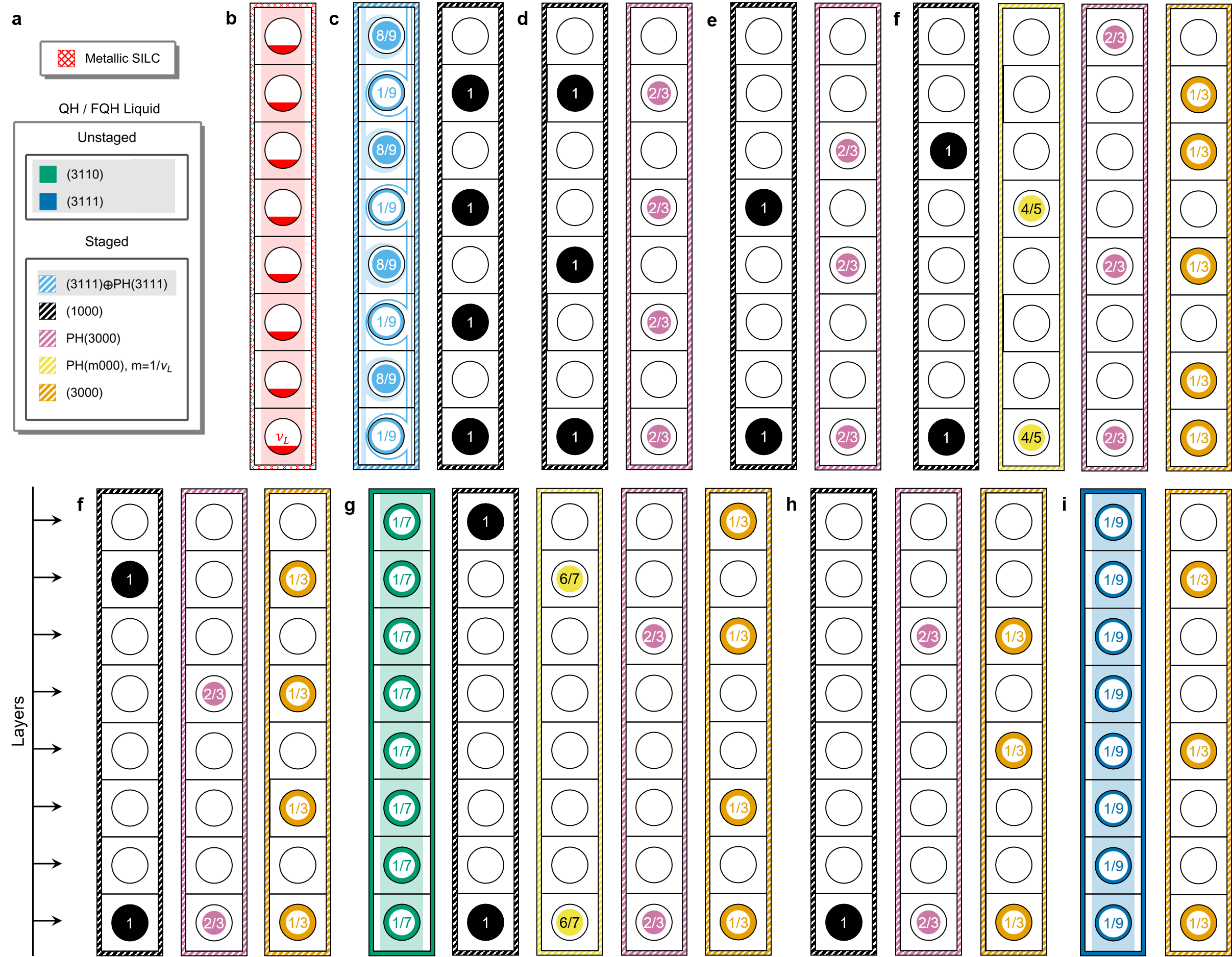}
\caption{{\bf Entanglement and staging patterns of transition metal dichalcogenides.} {\bf a,} Symmetry-breaking and entanglement patterns of metallic spontaneous interlayer coherent states with filling fraction per layer \(\nu_L\), and {\bf b,} quantum Hall states with \(\nu_L=1/2\), {\bf c,} \(\nu_L=1/3\), {\bf d,} \(\nu_L=1/4\), {\bf e,} \(\nu_L=1/5\), {\bf f,} \(\nu_L=1/6\), {\bf g,} \(\nu_L=1/7\), {\bf h,} \(\nu_L=1/8\), {\bf i,} \(\nu_L=1/9\) in Fig.~\ref{figS4}. Each state is indicated by its frame. {\bf j,} Phase labels. Interlayer-coherent fractional quantum Hall phases are highlighted in gray.}
\label{figS5}
\end{center}
\end{figure}

\paragraph{Entanglement and Staging Patterns} Although there is no valley ordering to consider in TMDs, they still exhibit some particular entanglement and staging patterns. ~\ref{figS5} provides full schematics of entanglement and staging patterns of all phases in ~\ref{figS4}. We note that these patterns are not dependent on the twist angle \(\theta\), for both alternating and helical twisted multilayer TMDs.

\clearpage

\supnotesection{Quasiparticle Charges and their Braiding Statistics in Generalized Halperin States}

\begin{figure}[H]
\begin{center}
\includegraphics[width=1\textwidth]{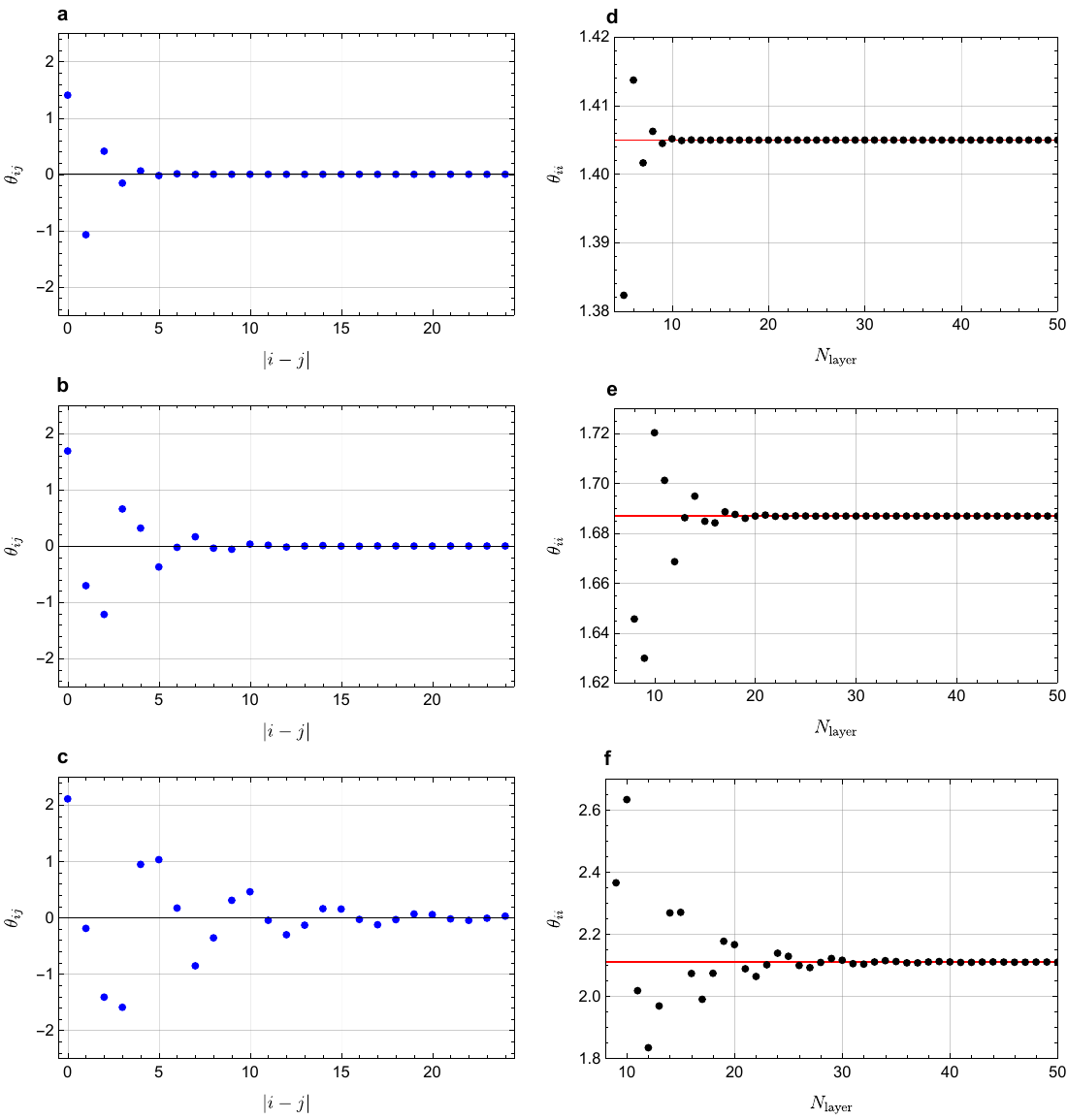}
\caption{{\bf Anyons with irrational statistics.} {\bf a,} Braiding phases \(\theta_{ij}\) of generalized Halperin liquids \((3100)\), {\bf b,} \((3110)\), and {\bf c,} \((3111)\) for various \(|i-j|\). {\bf d,} Self-statistical angles \(\theta_{ii}\) of \((3100)\), {\bf e,} \((3110)\), and {\bf f,} \((3111)\) liquids in finite \(N_\mathrm{layer}\)-layer systems for various \(N_\mathrm{layer}\) (black), and in the infinite-layer system (red).}
\label{figS6}
\end{center}
\end{figure}

We note that the generalized Halperin states with interlayer coherence, in particular, \((3100)\), \((3110)\), and \((3111)\) states, have rational quasiparticle charges with irrational braiding statistics. Prohibited in two-dimensions, the irrationality is intrinsically a three-dimensional property. Moreover, their statistical angles \(\theta_{ij}\) exponentially decay along \(|i-j|\) but are nontrivial (\ref{figS6}a--c). Such observation ensues the non-foliated property of the generalized Halperin states.

\paragraph{\((3100)\) State} The quasiparticle charge is \(q_i = -e/5\), and the phase angles of the braiding statistics  are 
\begin{equation}
\theta_{ii} = \frac{\pi}{\sqrt{5}},\qquad \theta_{ij} = 2\pi \times \frac{(-1)^{i-j}}{\sqrt{5}}\left(\frac{3+\sqrt{5}}{2}\right)^{-|i-j|} \quad(i\neq j).
\end{equation}
\(\theta_{ij}\) for various \(|i-j|\) are plotted in \ref{figS6}a.

\paragraph{\((3110)\) State} The quasiparticle charge is \(q_i = -e/7\), and the phase angles of the braiding statistics are
\begin{align}
\theta_{ii} &= \pi\sqrt{\frac{1}{7}+\frac{2}{3\sqrt{21}}}, \nonumber \\
\theta_{ij} & = 2\pi \times \Bigg[\left(\frac{i}{2\sqrt{3}} - \frac{1}{2}\sqrt{-\frac{1}{21}-\frac{2i}{21\sqrt{3}}}\right)\left(-\frac{1}{4} + \frac{i\sqrt{3}}{4} + \frac{1}{2}\sqrt{-\frac{9}{2}-\frac{i\sqrt{3}}{2}}\right)^{1+|i-j|} \nonumber \\
 & \qquad \qquad + \left(-\frac{i}{2\sqrt{3}} - \frac{1}{2}\sqrt{-\frac{1}{21}+\frac{2i}{21\sqrt{3}}}\right)\left(-\frac{1}{4} - \frac{i\sqrt{3}}{4} + \frac{1}{2}\sqrt{-\frac{9}{2}+\frac{i\sqrt{3}}{2}}\right)^{1+|i-j|}\Bigg] \quad(i\neq j).
\end{align}
\(\theta_{ij}\) for various \(|i-j|\) are plotted in \ref{figS6}b.

\paragraph{\((3111)\) State} The quasiparticle charge is \(q_i = -e/9\), and the phase angles of the braiding statistics are
\begin{align}\label{3111braidingstats}
\theta_{ii} &= \frac{\pi}{279}\left[31 + \sqrt[3]{124}\left(\sqrt[3]{3937 - 99\sqrt{93}} + \sqrt[3]{3937 + 99\sqrt{93}}\right)\right], \nonumber \\
\theta_{ij} & = 2\pi \times \Bigg[\frac{1}{279}\left(62 + \sqrt[3]{\frac{31(1147-117\sqrt{93}}{2}} + \sqrt[3]{\frac{31(1147+117\sqrt{93}}{2}}\right)\left(\sqrt[3]{\frac{-9+\sqrt{93}}{18}} - \sqrt[3]{\frac{9+\sqrt{93}}{18}}\right)^{|i-j|+2} \nonumber \\
& \qquad \qquad + \left(-\frac{2}{9}-\frac{\left(1+i\sqrt{3}\right)}{18\sqrt[3]{31^2}}\sqrt[3]{\frac{5549+603\sqrt{93}}{2}}+\frac{\left(1-i\sqrt{3}\right)}{18\sqrt[3]{31^2}}\sqrt[3]{\frac{-5549+603\sqrt{93}}{2}}\right) \nonumber \\
& \qquad \qquad \qquad \times \left(-\frac{1}{3} + \frac{\left(1-i\sqrt{3}\right)}{6}\sqrt[3]{\frac{29-3\sqrt{93}}{2}} + \frac{\left(1+i\sqrt{3}\right)}{6}\sqrt[3]{\frac{29+3\sqrt{93}}{2}}\right)^{|i-j|+2} \nonumber \\
& \qquad \qquad+ \left(-\frac{2}{9}-\frac{\left(1-i\sqrt{3}\right)}{18\sqrt[3]{31^2}}\sqrt[3]{\frac{5549+603\sqrt{93}}{2}}+\frac{\left(1+i\sqrt{3}\right)}{18\sqrt[3]{31^2}}\sqrt[3]{\frac{-5549+603\sqrt{93}}{2}}\right) \nonumber \\
& \qquad \qquad \qquad \times \left(-\frac{1}{3} + \frac{\left(1+i\sqrt{3}\right)}{6}\sqrt[3]{\frac{29-3\sqrt{93}}{2}} + \frac{\left(1-i\sqrt{3}\right)}{6}\sqrt[3]{\frac{29+3\sqrt{93}}{2}}\right)^{|i-j|+2}\Bigg]  \quad(i\neq j).
\end{align}
\(\theta_{ij}\) for various \(|i-j|\) are plotted in \ref{figS6}c.

\paragraph{Self-Statistics in Finite Systems} The self-statistics of anyons in the finite system of \(N_\mathrm{layer}\) layers converges to the irrational statistics of the infinite-layer systems as \(N_\mathrm{layer}\) increases (\ref{figS6}d--f). The convergence is faster as the coherence spans less. However, even the \((3111)\) state whose coherence spans up to third-nearest neigbors, the deviance from the irrational self-statistical angle is less than $0.1$ rad for the system with \(N_\mathrm{layer} > 20\). 

\clearpage

We remark that if the `seed' states exhibit rational quasiparticle charges with irrational braiding statistics, so do their product and particle-hole conjugate states. That is, \((3111)\oplus\mathrm{PH}(3111)\), \((3110)\oplus(3110)\), and \((3111)\oplus(3111)\) states in the main text (Figs.~2,3) host the anyons with rational charges but irrational braiding statistics.

\paragraph{Product States} Let \(X_1\) be a state whose quasiparticle is \(q_{(i,1)}\) and braiding phase is \(\theta_{(i,1)(j,1)}\) for each \(i,j\in I_1\). Similarly, let \(X_2\) be a state whose quasiparticle is \(q_{(i,2)}\) and braiding phase \(\theta_{(i,2)(j,2)}\) for each \(i,j\in I_2\). Then, the quasiparticle charge of the product state \(X_1 \oplus X_2\) is \(q_i\) for each \(i\in I\), where 
\begin{equation*}
	I = I_1 \oplus I_2 = \bigcup_{a=1}^2\{(x,a):x\in I_a\}.
\end{equation*} 
Moreover, the braiding phase is given by \(\theta_{ij}\) for each \(i,j\in I\), while \(\theta_{(x,1)(y,2)} = \theta_{(y,2)(x,1)} = 0\) for each \(x\in I_1\) and for each \(y \in I_2\).

\paragraph{Particle-Hole Conjugate States} Let \(X\) be a state whose quasiparticle is \(q_i\) and braiding phase is \(\theta_{ij}\) for each \(i,j\in I\). Then, the quasiparticle charge and the braiding phase of the particle-hole conjugate state \(\bar{X}\) is 
\begin{equation*}
	q_{(i,n)} = \begin{cases} q_i & \quad \text{if } n = 1\\
	-e & \quad \text{if } n = 2
	\end{cases},\quad
    \theta_{(i,n)(j,m)} = \begin{cases}
	-\theta_{ij} & \quad \text{if } n = m = 1 \\
	\pi & \quad \text{if } n = m = 2 \text{ and } i=j \\
	0 & \quad \text{otherwise}
	\end{cases}
\end{equation*}
for each \(i,j\in I\) and \(n,m\in\{1,2\}\).

\supnotesection{Energy Evaluation Schemes}

\paragraph{Estimation of the Coulomb Energy in Infinite-Layer Systems} We estimate the Coulomb energy of the infinite-layer system with given particle configurations on spheres as follows. For each layer, the positive uniform background charge is assumed to compensate the negative charge density of electrons. We note that this assumption is necessary for the convergence of the Coulomb energy per particle, \(E_{\mathrm{Coulomb}}/N_{\mathrm{tot}}\) for infinite-layer systems. Moreover, we simply omit the layer and valley indices of the spinor variables, and let \((u_i, v_i)\) be on layer \(l\), and \(u_i', v_i'\) be on layer \(l'\) for any possible \(i\). The Coulomb interaction energy consists of three parts:
\begin{equation}\label{Coulombsum}
    E_{\mathrm{Coulomb}} = E_\mathrm{el-el} + E_\mathrm{el-bkg} + E_\mathrm{bkg-bkg},
\end{equation}
where
\begin{align}
    E_\mathrm{el-el}=\frac{e^2}{2\epsilon R} \sum_{l,l'} \Bigg[& \sum_{\substack{i,j \\ (l,i)\neq(l',j)}} \Bigg(\frac{1}{\sqrt{|u_iv_j'-v_iu_j'|^2+\alpha^2|l-l'|^2}}\Bigg) \nonumber \\
    &+\sum_{i,j} \sum_{m\in\mathbb{Z}\backslash\{0\}} \Bigg(\frac{1}{\sqrt{|u_iv_j'-v_iu_j'|^2+\alpha^2|(l-l')+mN_L|^2}}\Bigg)\Bigg].
\end{align}
is the electron-electron interaction energy, 
\begin{equation}
    E_\mathrm{el-bkg}=-\frac{e^2}{\epsilon R}\sum_{l,l'} \Bigg[N_l N_{l'}\sum_{m\in\mathbb{Z}} \Bigg(\sqrt{1+\alpha^2\big|(l-l')+mN_L\big|^2}-\alpha\big|(l-l')+mN_L\big|\Bigg)\Bigg].
\end{equation}
is the electron-background interaction energy, and
\begin{equation} \label{intbetlayers}
    E_\mathrm{bkg-bkg}=\frac{e^2}{2\epsilon R}\sum_{l,l'} \Bigg[N_l N_{l'}\sum_{m\in\mathbb{Z}} \Bigg(\sqrt{1+\alpha^2\big|(l-l')+mN_L\big|^2}-\alpha\big|(l-l')+mN_L\big|\Bigg)\Bigg].
\end{equation}
is the background-background interaction energy, while \(N_l\) is the number of particles in layer \(l\), \(\alpha = d/2R\), \(R\) is the radius of the Haldane sphere, and \(N_L\) is the number of layers in the fundamental cell. For each \(1\leq l'< l\leq N_L\) and for each \(m\in\mathbb{N}_0\), we define 
\begin{equation}
    f_m^{(l)} =  -\frac{2N_l}{N_l-1} \sum_{i<j} a_m^{(l),(i,j)}, \quad f_m^{(l,l')} = -2\sum_{i,j} \sum_{\xi=\pm} a_m^{(l,l',\xi),(i,j)},
\end{equation}
where 
\begin{align}
    a_0^{(l),(i,j)} & = 1 - \frac{N_l - 1}{2N_l}\frac{1}{|u_iv_j-v_iu_j|},  \nonumber \\
        a_m^{(l),(i,j)} & = \sqrt{1+\beta^2m^2}-\beta m - \frac{1}{2N_l}\left(\frac{1}{\beta m} + \frac{N_l - 1}{\sqrt{|u_iv_j-v_iu_j|^2+\beta^2m^2}}\right) \quad(m\neq 0),\nonumber\\
    a_0^{(l,l',+),(i,j)} & = \sqrt{1+\beta^2 k_{l-l'}^2} -\beta k_{l-l'} - \frac{1}{2\sqrt{|u_iv_j'-v_iu_j'|^2+\beta^2 k_{l-l'}^2}}, \qquad a_0^{(l,l',-),(i,j)} = 0,  \nonumber \\
    a_m^{(l,l',\pm),(i,j)} & = \sqrt{1+\beta^2(m\pm k_{l-l'})^2} -\beta (m\pm k_{l-l'}) - \frac{1}{2\sqrt{|u_iv_j'-v_iu_j'|^2+\beta^2(m\pm k_{l-l'})^2}}\quad(m\neq 0),
\end{align}
and \(\beta = \alpha N_L\), to rewrite Eq.~\eqref{Coulombsum} as
\begin{equation}
    E_{\mathrm{Coulomb}} = \frac{e^2}{2\epsilon R} \left[\sum_{l} \left(f_0^{(l)}+ 2\sum_{m\in\mathbb{N}} f_m^{(l)}\right) + \sum_{l>l'}\left(f_0^{(l,l')}+\sum_{m\in\mathbb{N}} f_m^{(l,l')}\right)\right].
\end{equation}
Assuming \(|\beta m| \gg 1\), the binomial approximation gives 
\begin{equation}
    a_m^{(l),(i,j)} \approx \frac{N_l-1}{4N_l}\frac{1}{\beta^3 m^3}|u_iv_j-v_iu_j|^2, \quad
    a_m^{(l,l',\pm),(i,j)} \approx \frac{1}{4}\frac{1}{\beta^3(m\pm k_{l-l'})^3}|u_iv_j'-v_iu_j'|^2,
\end{equation}
resulting in 
\begin{equation}
    f_m^{(l)} \approx -\frac{1}{2\beta^3}\frac{1}{ m^3} \left(\sum_{i<j}\left|u_iv_j-v_iu_j\right|^2\right),\  
    f_m^{(l,l')} \approx -\frac{1}{2\beta^3}\left(\frac{1}{(m+k_{l-l'})^3}+\frac{1}{(m-k_{l-l'})^3}\right) \left(\sum_{i,j}\left|u_iv_j'-v_iu_j'\right|^2\right).
\end{equation}
We remark that for any \(-1 < k < 1\) and for any \(M\in\mathbb{N}\),
\begin{equation*}
    \sum_{m=M}^\infty \frac{1}{(m+k)^3} = \sum_{m=0}^\infty \frac{1}{(m+M+k)^3} = \zeta(3,M+k) \text{ (Hurwitz zeta function)}.
\end{equation*}
Then, choosing a cutoff \(M \gg 1/\beta\), the below approximation is valid:
\begin{equation} \label{approxCoulomb}
    E_{\mathrm{Coulomb}} \approx \frac{e^2}{2\epsilon R} \Bigg[\sum_{l} \left(f_0^{(l)} + f_\infty^{(l)} + 2\sum_{m=1}^{M-1} f_m^{(l)}\right) + \sum_{l>l'}\left(f_0^{(l,l')} + f_\infty^{(l,l')}+\sum_{m=1}^{M-1} f_m^{(l,l')}\right)\Bigg],
\end{equation}
where 
\begin{align}
    f_\infty^{(l)} & = -\frac{1}{\beta^3}\zeta(3,M)\left(\sum_{i<j}\left|u_iv_j-v_iu_j\right|^2\right) \quad (1\leq l\leq N_L), \nonumber \\
    f_\infty^{(l,l')} & = -\frac{1}{2}\frac{1}{\beta^3}\bigg(\zeta(3,M+k_{l-l'})+\zeta(3,M-k_{l-l'})\bigg)\left(\sum_{i,j}\left|u_iv_j'-v_iu_j'\right|^2\right) \quad (1\leq l' < l\leq N_L).
\end{align}
We adopt Eq.~\ref{approxCoulomb} to compute the Coulomb energy from given particle configurations.

\paragraph{Monte Carlo Method} Energies of liquid and crystal states are determined from Monte Carlo method. For each state and each \(d/l_B\), we use Metropolis-Hastings algorithm to sample the particle configurations. The overall procedure is described in the Algorithm~\ref{mcalgo}. In particular, the candidate is chosen to be a random point in the \(\delta\)-neighborhood of the position of our interest for each run, while \(\delta\) is determined from interpolations of acceptance rates of the prerun data.

\normalem 
\begin{algorithm*}[H]
\DontPrintSemicolon
\Fn{\Sweep{$\{x_1',x_2',\cdots,x_N'\}$}}{
\Input{Particle configuration $\{x_1',x_2',\cdots,x_N'\}$}
\Output{Particle configuration $\{x_1,x_2,\cdots,x_N\}$}
$\{x_1,x_2,\cdots,x_N\} \leftarrow \{x_1',x_2',\cdots,x_N'\}$\;
\For{$i=1$ to $N$}{
$x\leftarrow$ a random candidate to replace $x_i$\;
$\displaystyle a \leftarrow\frac{|\Psi(\{x_1, x_2,\cdots, x_{i-1}, x_i, x_{i+1}, \cdots, x_{N-1}, x_N\})|^2}{|\Psi(\{x_1, x_2,\cdots, x_{i-1}, x, x_{i+1}, \cdots, x_{N-1}, x_N\})|^2}$\; \Comment{$|\Psi(X)|^2$ is the probability amplitude of a wavefunction $\Psi$ with the particle configuration $X$.}
$r\leftarrow$ a random number from $[0,1)$\;
\If{$a\geq r$}{
$x_i\leftarrow x$\;
}
}
\Return{$\{x_1,x_2,\cdots,x_N\}$}
}
\;
\Fn{\Metropolis{$\bar{N}, N_L, M, N_{\mathrm{sweep}},N_{\mathrm{burn-in}},N_{\mathrm{sample}}$}}{
\Input{Average number of particles per layer $\bar{N}$, number of layers in the fundamental cell $N_L$, cutoff $M$, number of sweeps $N_{\mathrm{sweep}}$, number of draws discarded $N_{\mathrm{burn-in}}$, number of samples collected $N_{\mathrm{sample}}$}
\Output{Total energies $\{E_1,E_2,\cdots,E_{N_{\mathrm{sample}}}\}$}
$\{E_1,E_2,\cdots,E_{N_{\mathrm{sample}}}\}$\;
$\{x_1,x_2,\cdots,x_{\bar{N}\cdot N_L}\}\leftarrow$ a particle configuration whose particles are at the random positions on the sphere\;
\For{$i=1$ to $N_{\mathrm{burn-in}}$}{
\For{$j=1$ to $N_{\mathrm{sweep}}$}{
$\{x_1,x_2,\cdots,x_{\bar{N}\cdot N_L}\}\leftarrow$\Sweep{$\{x_1,x_2,\cdots,x_{\bar{N}\cdot N_L}\}$}\;
}
}
\For{$i=1$ to $N_{\mathrm{sample}}$}{
\For{$j=1$ to $N_{\mathrm{sweep}}$}{
$\{x_1,x_2,\cdots,x_{\bar{N}\cdot N_L}\}\leftarrow$\Sweep{$\{x_1,x_2,\cdots,x_{\bar{N}\cdot N_L}\}$}\;
}
$E_i\leftarrow$ total energy of the system with particle configuration $\{x_1,x_2,\cdots,x_{\bar{N}\cdot N_L}\}$ computed with the cutoff $M$
}
\Return{$\{E_1,E_2,\cdots,E_{N_{\mathrm{sample}}}\}$}
}
\caption{Metropolis-Hastings sampling \cite{NatComm2025_Kim}.}\label{mcalgo}
\end{algorithm*}
\ULforem 

Assuming that every layer has the same amount of particles, we choose \(\bar{N} = 48\), \(N_\mathrm{burn-in} = 10^3\), and  \(N_\mathrm{sample} = 10^6\), while \(N_\mathrm{sweep}\) is determined from the integrated autocorrelation time computed from the pre-run data. If the state is valley-coherent, \(\bar{N}\) is equally divided between the valleys. Moreover, letting the radius of the Haldane sphere be \(R\), \(\alpha=d/2R\), and \(\beta = \alpha N_L\), we choose the cutoff \(M = M_{\mathrm{eff}}/\alpha\) from the relative cutoff parameter \(M_{\mathrm{eff}}\) so that \(M >> 1/\beta\) is satisfied. List of \(M_{\mathrm{eff}}\) is given in \ref{TableS9}. 

\begin{table}[htbp]
    \centering
    \begin{tabular}{c || c | c | c | c | c | c | c}
    \hline
         \(d/l_B\) & 0.05 & 0.1 & 0.15 & 0.2 & 0.3 & 0.4 & \(\geq 0.6 \) \\
         \hline\hline
         \(M_\mathrm{eff}\) & 0.9795 & 1.069 & 1.175 & 1.275 & 1.5 & 1.75 & 2 \\
    \hline
    \end{tabular}
    \caption{List of relative cutoff parameter \(M_{\mathrm{eff}}\) adopted for each \(d/l_B\). Each \(M_{\mathrm{eff}}\) is determined from the benchmark calculations for decoupled stacks of \(1/3\) Laughlin states to ensure convergence.}
    \label{TableS9}
\end{table}

The Coulomb energy is estimated from the average of total energies from the Metropolis-Hastings sampling,
\begin{equation}
    E_\mathrm{Coulomb} = \sqrt{\frac{N_\phi \bar{\nu}}{\bar{N}}} \cdot \frac{1}{N_\mathrm{sample}} \sum_{i=1}^{N_\mathrm{sample}} E_i,
\end{equation}
with the square root factor for the density correction to weaken the energy dependence on the number of particles \cite{Book_Jain, NatComm2025_Kim}. Computing the Coulomb energy per particle \(E_{\mathrm{Coulomb}}/N_{\mathrm{tot}}\) for various \(N_L\), we obtain the thermodynamic-limit value by least-squares extrapolation of \(E_{\mathrm{Coulomb}}/N_{\mathrm{tot}}\) as a function of \(1/N_L\) to \(N_L\rightarrow\infty\), from which the phase diagrams are determined.

\paragraph{Mean-Field Theory with Phenomenological Short-Range Interactions} Since the phenomenological short-range interactions are elaborated by the mean-field approach~\cite{NatComm2017_Hunt, NanoLett2023_Kim, NatComm2025_Kim, NanoLett2026_Kim, PRB2006_Alicea, PRB2012_Kharitonov, PRL2012_Kharitonov, PRL2014_Sodemann}, we adopt the same framework to determine their energy contribution. In particular, the mean-field Hamiltonian is given by
\begin{align}\label{meanfield}
	H_V^{\mathrm{MF}} =  \sum_{l}  \int d\mathbf{r} \  &V_1 \Big(\langle n_{l,K}(\mathbf{r}) \rangle n_{l,-K}(\mathbf{r}) +  n_{l,K}(\mathbf{r}) \langle n_{l,-K}(\mathbf{r}) \rangle - \langle n_{l,K}(\mathbf{r}) \rangle \langle n_{l,-K}(\mathbf{r}) \rangle \Big) \nonumber \\ + \sum_\sigma \Big[
	&V_2 \Big( \langle n_{l,\sigma}(\mathbf{r}) \rangle n_{l+1,\sigma}(\mathbf{r}) + n_{l,\sigma}(\mathbf{r}) \langle n_{l+1,\sigma}(\mathbf{r}) \rangle - \langle n_{l,\sigma}(\mathbf{r}) \rangle \langle n_{l+1,\sigma}(\mathbf{r}) \rangle \Big)
	\nonumber \\ + & V_3 \Big(\langle n_{l,\sigma}(\mathbf{r}) \rangle  n_{l+1,-\sigma}(\mathbf{r}) +  n_{l,\sigma}(\mathbf{r}) \rangle \langle n_{l+1,-\sigma}(\mathbf{r}) - \langle n_{l,\sigma}(\mathbf{r}) \rangle \langle n_{l+1,-\sigma}(\mathbf{r}) \rangle \Big) \Big],
\end{align}
where \(n_{l,\sigma}(\mathbf{r}) = \hat{\psi}^\dagger_{l,\sigma}(\mathbf{r}) \hat{\psi}_{l,\sigma}(\mathbf{r})\) for each layer \(l\) and valley \(\sigma\). Summing up the Coulomb energy, the tunneling energy, and the energy evaluated from Eq.~\eqref{meanfield}, we calculate total energies of the candidates, from which we find the ground states. Moreover, under the same physical intuition, we simply fix the momentum difference for every consecutive pairs of occupied layers to be maximum for the staged states.

In most cases, the above strategy successfully determines the ground state. However, there are some exceptions where the liquid ground state turns out to be unphysical due to its infinite degeneracy. Despite its rigid topology, the phenomenological short-range interaction energy remains the same for all possible valley textures. We note that the mean-field method overestimates the phenomenological interaction energy of the liquid states with interlayer or intervalley correlations, which means the state with higher phenomenological interaction energy between the correlated layers and valleys actually has a lower phenomenological interaction energy. Following this logic, we determine the valley texture of the ground state if the exception occurs.

\bibliography{supprefs.bib}